\documentclass[aps,pra,showpacs,floats,twocolumn,unsortedaddress,floatfix,footinbib]{revtex4-1}
\usepackage{amsmath,amsfonts,amssymb,amsthm,mathtools}
\usepackage{mathrsfs,stmaryrd,bm,wasysym}
\usepackage{accents}
\usepackage{enumitem}
\usepackage{graphicx}
\usepackage{hyperref}
\usepackage{breakurl}
\usepackage{color}
\usepackage{tikz-cd}

\numberwithin{equation}{section}
\newcounter{dummy}
\makeatletter
\newcommand\myitem[1][]{\item[#1]\refstepcounter{dummy}\def\@currentlabel{#1}}
\makeatother

\theoremstyle{plain}
\newtheorem{thm}{Theorem}[section]
\newtheorem{lem}[thm]{Lemma}
\newtheorem{cor}[thm]{Corollary}
\newtheorem{prop}[thm]{Proposition}

\theoremstyle{definition}
\newtheorem{defn}{Definition}[section]

\theoremstyle{remark}

\let\origmaketitle\maketitle
\def\maketitle{
  \begingroup
  \def\uppercasenonmath##1{} 
  \let\MakeUppercase\relax 
  \origmaketitle
  \endgroup
}

\newcommand{\defeq}{ {\kern 0.2em}:{\kern -0.5em}={\kern 0.2 em} }  
\newcommand{\eqdef}{ {\kern 0.2em}={\kern -0.5em}:{\kern 0.2 em} }  
\newcommand{\setof}[2]{\left\{ #1 \;:\; #2 \right\}}  

\newcommand{\Arr}[3]{{#2}\colon {#1} \rightarrow {#3}} 

\DeclareMathOperator{\linspan}{span}  

\newcommand{\Nat}{{\mathbb N}}     
\newcommand{\Rat}{{\mathbb Q}}     
\newcommand{\Real}{{\mathbb R}}    
\newcommand{\Cmplx}{{\mathbb C}}   

\newcommand{\dom}{\mathrm{dom}\,}

\newcommand{\pair}[2]{\left\langle {#1} \,,\, {#2} \right\rangle}

\DeclareMathOperator{\Tr}{\mathrm{Tr}}

\newcommand{\tgt}[1]{{#1}^{\scriptscriptstyle\odot}}

\DeclareMathOperator\dens{\mathsf{dens}}
\newcommand{\wh}[1]{ \widehat{#1} }

\newcommand{\xs}{{\Delta}}
\newcommand{\zero}{{\mathscr Z}}

\newcommand{\slct}[1]
{\left[ {#1} \right] }
\newcommand{\tinyHK}{\textrm{\tiny{HK}}}
\DeclareMathOperator{\KS}{\mathsf{KS}}
\DeclareMathOperator{\Rsd}{\mathsf{R}}

\newcommand{\ktiny}{\prec{\kern -0.8em}\prec}

\newcommand{\diff}{{\mathsf D}}
\newcommand{\ldiff}{\underline\diff}
\newcommand{\udiff}{\overline\diff}
\newcommand{\subdiff}{\ldiff}
\newcommand{\supdiff}{\udiff}

\newcommand{\Obs}{\mathsf{Obs}}
\newcommand{\Sts}{\mathsf{Sts}}

\newcommand{\NN}{\mathcal{N}}
\newcommand{\MM}{\mathcal{M}}
\newcommand{\HH}{\mathcal{H}}

\DeclareMathOperator{\Num}{\mathrm{Num}}
\newcommand{\SX}{\mathscr{X}}
\newcommand{\SY}{\mathscr{Y}}
\newcommand{\SD}{\mathscr{D}}

\newcommand{\SV}{\mathscr{V}}

\newcommand{\inc}{\mathrm{inc}}
\DeclareMathOperator{\vecsp}{\mathrm{Vec}}

\newcommand{\defined}{iff\ } 

\begin{document}

\title{
Kohn-Sham computation and the bivariate view of density functional theory
}
\author{Paul~E.~Lammert}
\email{lammert@psu.edu}
\affiliation{Department of Physics, 104B Davey Lab \\ 
Pennsylvania State University \\ University Park, PA 16802-6300}
\date{Sept. 7, 2023}
\begin{abstract}
Informed by an abstraction of Kohn-Sham computation called a KS machine,
a functional analytic perspective is developed on mathematical aspects of
density functional theory. A natural semantics for the machine is bivariate,
consisting of a sequence of potentials paired with a ground density.
Although the question of when the KS machine can converge to a solution
(where the potential component matches a designated target) is not
resolved here, a number of related ones are. For instance:
Can the machine progress toward a solution? Barring presumably
exceptional circumstances, yes in an energetic sense, but using a
potential-mixing scheme rather than the usual density-mixing variety.
Are energetic and function space distance notions of proximity-to-solution
commensurate? Yes, to a significant degree.
If the potential components of a sequence of ground pairs converges
to a target density, do the density components cluster on ground
densities thereof? Yes, barring particle number drifting to infinity.
\end{abstract}
\maketitle

\section{Introduction}

Density functional theory (DFT)
has developed into a ubiquitous tool in physics, chemistry, materials science, and 
beyond\cite{Hohenberg+Kohn,Parr+Yang,Dreizler+Gross,Koch+Holthausen,Capelle06,Burke-12},
overwhelmingly in the specific form of Kohn-Sham\cite{Kohn+Sham} (KS) computation.
The two distinguishing features of KS computation are (i) a splitting of the intrinsic
energy functional into noninteracting, Hartree, and exchange-correlation contributions,
and (ii) an idiosyncratic procedure of iterating to so-called self-consistency.
Meanwhile, the functional analytic approach initiated by Lieb\cite{Lieb83}
has had little\cite{Lieb+Oxford-81,Laestadius+18-JCP,Laestadius+19-JCTC,Penz+19-PRL,Penz+19-erratum} to say about these things. Working in the functional analytic tradition,
this paper aims both at filling that gap, and at developing a more physical
interpretation of KS computation. Pursuit of these goals is synergetic,
as the following sketch of major themes shows.

\subsection{Appetizer}

What is the physical interpretation of intermediate stages of a KS computation,
i.e., before self-consistency is achieved?
The course of the computation can be cast (sec. \ref{sec:ground-pairs})
as a sequence of \textit{ground pairs} ---
pairs consisting of a potential and a corresponding (interacting) ground density.
This transparent framing is a promising basis for both theoretical analysis
and algorithmic development.
Thinking of potential and density simultaneously variable, we have moved
into a bivariate perspective. The action takes place in the \emph{product}
of potential and density space.

For an iterative procedure to find a ground density of a given (\textit{target})
potential, it first needs to \emph{make progress} toward that goal from one iteration
to the next.
A scheme using just the usual Kohn-Sham computational resources, 
is described (sec. \ref{sec:feasible-strategy}), which makes progress
in the sense of finding a density with lower energy in the target potential,
barring exceptional circumstances (hitting a potential with a degenerate
ground state or none at all, lack of exchange-correlation potential).
The proposed scheme involves potential mixing, in contrast to the usual density-mixing ones.

Such progress is far short of convergence. However, as already noted, we can
cast all KS computations as sequences of ground pairs.
Suppose, optimistically, that we have such a sequence $(v_n,\rho_n)$
for which $v_n$ converges to the target potential.
Does it follow that the densities $\rho_n$ converges to a target ground density? 
The pleasant answer (Sec.~\ref{sec:RCIII}) is that the sequence of densities $(\rho_n)$
clusters with respect to $L^1$ metric at target ground densities, as long
as it does not have nonzero particle number drifting to infinity.
If the target ground density is unique, this means the sequence converges.
In motto form: look after the potential and the density will take care of
itself. With the finding about a potential-mixing scheme, this supports the
idea that current ways of thinking are too density-centric.
Some trends in computational practice, such as the use of hybrid functionals\cite{Becke-93}
are in harmony with this thought.

It is known, since Lieb's seminal work\cite{Lieb83}, that the intrinsic energy
functional $F$
(a.k.a. Levy or Levy-Lieb functional, see section~\ref{sec:intrinsic-energy}) is not continuous.
Indeed it is unbounded above on every neighborhood with respect to natural topologies.
Should the practitioner be worried about that?
A surprisingly encouraging answer emerges (Secs.~\ref{sec:RCI} and \ref{sec:RCII}).
Restricted to the set of ground pairs, $F$ is {continuous}
(with respect to \emph{product} topology, that is).
Living on this subset of ground pairs within the product space, 
KS computation is, in a sense, insulated from the discontinuity.

\subsection{Outline}

Section \ref{sec:QM-to-DFT} traces the reduction of quantum mechanics to
density functional theory, characterizing DFT as an observable/state theory.
This primitive physical framework must be the touchstone for all
mathematical, in particular topological, refinement.
Section \ref{sec:derivatives} presents a version of unilateral functional differentiation
for real-valued functions.
  To avoid explicitly introducing topological considerations at this stage,
  derivatives are defined in an unusual way, relative to a dual pair.
Section \ref{sec:KS-machines} analyzes the basic operations of KS computation and
the ways they can be combined, and abstracts these resources in the form of
a \emph{Kohn-Sham machine}.
The bivariate view and the \textit{excess energy} ($\xs$) make their appearance here.
For an exact functional, $\xs(v,\rho)$ is the lowest energy achievable by
a quantum state of density $\rho$ in potential $v$, relative to the ground energy,
thus quantifying the ``mismatch'' of $v$ and $\rho$.
This is a very natural function to work with in the bivariate view.
Section \ref{sec:progress} examines the possibility of guaranteed
progress in the sense of reducing $\xs(\tgt{v},\rho)$, where $\tgt{v}$ is the
target potential, verifiable by the resources of a KS machine.
A proposed scheme is argued to usually (\textit{vide infra}
for the meaning of this) be able to progress. It is a potential-mixing scheme
in contrast to the usual density-mixing schemes, which we are unable to
meaningfully analyze.
A general discussion of semimetrics and metrics in Section \ref{sec:topology}
prepares the way to bring topology into consideration.
This is essential for discussing convergence and approximation in density-potential space.
Observable-state duality is the basis for the development here.
Because we take the question of what kind of metrical structure is appropriate on these
spaces to be a physical question, the mathematics we are pushed into is possibly
more sophisticated than might seem quite natural from a purely mathematical point of view,
which would simply declare densities to be living in a particular Banach space and get
on with the theorems.
Sections \ref{sec:RCI} and \ref{sec:RCII} are concerned with the character of the
main functionals, intrinsic energy, ground energy (Section \ref{sec:to-DFT}),
and excess energy (Section \ref{sec:excess-energy}) on the product space $\SV\times\SD$
of potential-density pairs.
One of the more notable findings is that, although the intrinsic energy $F$
is unbounded above on every neighborhood, it's degree of discontinuity in a certain sense
is less and less as we consider it in restriction to a subspace of smaller and smaller
excess energy. 
Section \ref{sec:near-vs-nearly} 
examines how proximity in density-potential space to a ground pair compares to small
excess energy, showing that a pair of low excess energy is close to a ground pair,
and slightly perturbing the potential component of a ground pair increases the
excess energy only a little. 
Finally, section \ref{sec:RCIII} shows that convergence to $v$ of the potential
components of a sequence of ground pairs guarantees that the density
components accumulate on ground densities of $v$ as long as particle number
does not drift to infinity.
Throughout the paper, axioms are introduced one-by-one, verified on the
standard interpretation, and their consequences traced.
This helps to make their specific significance clearer.
Interludes serve to motivate steps of the development.

The reader should not hesitate to skip proofs and demonstrations
on a first reading.
The short \textit{Interlude} sections are intended as
guideposts, indicating where the development is going and why.
For readers wishing a modest-length introduction to mainstream DFT,
Ref. \cite{Capelle06} is suggested.

\subsection{Notational notes}

Parentheses are used for pairs, e.g., $(v,\rho)$, and also for sequences,
e.g., $(x_n)_{n\in\Nat}$.
However, the index is usually obvious, so we can write $((v_n,\rho_n))$,
for a sequence of potential-density pairs, or even just $(v_n,\rho_n)$
since sequences of this specific type are ubiquitous here.
Limit inferior is denoted $\liminf$ or $\varliminf$, correspondingly,
limit superior by $\limsup$ or $\varlimsup$.
The abbreviation ``iff'' is used for ``if and only if''.
Functions on a function space are sometimes called functionals, sometimes not.

\section{From QM to DFT}\label{sec:QM-to-DFT}

This section sketches a view of DFT as a common sort of state/observable theory.
Classical as well as quantum theories can be framed this way.
However, we develop the theme only to an extent which can reasonably ground the subsequent
development, obtaining DFT proper by a contraction of the full quantum mechanical description
of an $\NN$-particle system.
A motto for this section is: density is not an observable.

The viewpoint of this section on the relation between DFT and QM is analogous
to that between thermodynamics and statistical mechanics.
Thermodynamics is \emph{a} theory, in the sense of giving an autonomous
description of certain aspects of the world and having its own proper
vocabulary and concepts. It is weak in the sense that it does not have
the resources to compute equations of state or free energy functions.
For that, one relies on statistical mechanics. However, thermodynamics
proper imposes constraints, for instance, a free energy must be convex in
certain of its variables, and concave in the others.
One of the aims is to formulate DFT as a theory in an analogous way.

\subsection{general quantum mechanics}
\label{sec:general-QM-example}

The setting for general quantum mechanics is a Hilbert space ${\mathcal H}$.
Observables ($\Obs$) are represented by bounded hermitian linear operators on $\HH$:
\begin{equation}
\Obs = {\mathcal B}_{sa}({\mathcal H}).
\end{equation}
States ($\Sts$) are represented by normalized, positive, trace-class operators:
\begin{equation}
\Sts = \mathcal{B}^1_{+}({\mathcal H}) \subset \mathcal{B}^1({\mathcal H}).
\end{equation}
Finally, there is a canonical pairing between observables and states given by
\begin{equation}
\pair{A}{\Gamma} = \Tr \Gamma A.
\end{equation}
This represents the expectation value of observable $A$ in the state $\Gamma$.
The notation on the LHS may seem gratuitous, however, it represents a general idea of
pairing observables and states
which may have different operational formulas (RHS) in different contexts.
This occurs in particular for DFT.
Pointy brackets are also a common notation in functional analysis for dual pairings
of vector spaces.
$\Obs$ is a vector space over $\Real$. $\Sts$ is not a vector space,
but it is identified as a subset of the vector space of trace-class
operators. Moreover, $\mathcal{B}^1(\HH)$ is the linear span of $\Sts$,
denoted $\linspan\Sts$.
The pairing naturally extends from a mapping $\Obs\times\Sts \rightarrow \Real$
to a mapping $\Obs\times\linspan\Sts \rightarrow \Real$, which is
\textit{bilinear}. In this sense, we can say that our observables are linear.

The more specific context that interests us is a system of $\NN$ identical
particles in three-dimensional space $\Real^3$, or more generally,
a three-dimensional riemannian manifold $\MM$.
The case of a three-torus shows that the more general situation is of
genuine interest.
For a single particle on $\MM$, the relevant Hilbert space is
$\HH_1 = L^2(\MM)$ if it is spinless, $\HH_1 = L^2(\MM)\otimes\Cmplx^2$ if
spin-${1}/{2}$. For the $\NN$-particle system, $\HH$ is the
symmetrized (bosons) or antisymmetrized (fermions) $\NN$-fold tensor
product of $\HH_1$. Everything we do will be valid for both cases.

\subsection{Function spaces defined by integrability conditions}

This subsection is not called on until (\ref{eq:Obs_1}), but placed here to
minimize disruption of the flow.

For measurable functions, we use the following standard notation for $1\le p < \infty$:
\begin{equation}
  \label{eq:Lp}
\|f\|_p \defeq \left(\int |f|^p\, dx\right)^{{1}/{p}}  \in [0,\infty].
\end{equation}
Actually, to be accurate, $f$ above should be considered an equivalence class of functions,
any two of which differ only on a set of measure zero. However, it is common to gloss over
the distinction, and we will follow that custom.
In addition, we define $\|f\|_\infty$ in $[0,\infty]$ to be the largest number such that
$\setof{x}{|f(x)| > \|f\|_\infty}$ has measure zero. For a bounded continuous function,
this is just the maximum, but more generally we again must accomodate measure-zero
exceptional sets. 

Now, we define the \textit{vector spaces}
\begin{equation}
\mathcal{L}^p(\mathcal{M}) = \setof{\text{measurable } f}{\|f\|_p < \infty}.
\end{equation}
$L^p(\mathcal{M})$ is $\mathcal{L}^p(\mathcal{M})$, equipped with $\|\cdot\|_p$ as
a norm.
At this stage of the development, we are using $\|\cdot\|_p$ only as a
selection mechanism. That is, there is no defined distance between members of
$\mathcal{L}^p(\mathcal{M})$.
Topological considerations (norms, seminorms and so forth) are deferred to
Section \ref{sec:topology}.

We need spaces a little more complicated than the pure $\mathcal{L}^p(\mathcal{M})$.
The intersections $\mathcal{L}^p(\mathcal{M}) \cap \mathcal{L}^q(\mathcal{M})$.
of the two spaces $\mathcal{L}^p(\mathcal{M})$ and $\mathcal{L}^q(\mathcal{M})$
is again a vector space, as is the sum
$\mathcal{L}^p(\mathcal{M}) + \mathcal{L}^q(\mathcal{M})$, consisting of all sums
of a function from each of the summand spaces.

\subsection{DFT}\label{sec:to-DFT}

Our development of DFT begins with a contraction of
the general QM observables, although we shall later expand to a set
which is neither subset nor superset of ${\mathcal B}({\mathcal H})$.

\subsubsection{Contracting QM}

We put subscripts on $\Obs$ and $\Sts$ to help avoid confusion, as there will be
more than one set.
Start with
\begin{equation}
  \nonumber
\Obs_0 \defeq \setof{\Num(U)}{U \text{ open in }\MM }. 
\end{equation}
Here, $\Num(U)$ is the operator reporting the number of particles in the set $U$.
(Choosing to start specifically with open sets is a somewhat arbitrary choice.)
Appealing to well-known facts about QM, the map
\begin{equation}
  \nonumber
U \mapsto \pair{\Num(U)}{\Gamma}  
\end{equation}
extends to a Borel measure, which is, moreover, absolutely continuous
with respect to Lebesgue measure (Fubini is helpful here).
This implies that there is an integrable function 
$\Arr{\MM}{\rho}{\Real}$ such that for any Lebesgue-measurable set $U$,
\begin{equation}
  \nonumber
  \pair{\Num(U)}{\Gamma}
  = \int_A \rho(x) \, dx.
\end{equation}
The measure theory just deployed is no cause for anxiety.
The main point is that, while in some other contexts
(e.g., classical statistical mechanical) we might want to consider Dirac measures,
the underlying QM precludes that here.
We give the QM-state-to-density mapping the name $\dens$.
Then, $\rho$ in the preceding integral is $\dens\Gamma$.

Where there is a measure, there is an integral. Indeed, we can write the
preceding formula as an integral of the indicator function $1(A)$, equal to
one on $A$, zero elsewhere:
\begin{equation}
  \nonumber
  \pair{\Num(U)}{\Gamma}
  = \int 1(U) (\dens\Gamma)(x) \, dx.
\end{equation}
This extends to \textit{some} measurable functions as 
\begin{equation}
  \pair{\Num(f)}{\Gamma}
  = \int f(x) (\dens\Gamma)(x) \, dx
  \label{eq:integral-f-Gamma}
\end{equation}
in the usual way, i.e., approximating $f$ by a linear combination of
indicator functions of sets. However, which functions $f$ are legitimate here?
If we are dealing with the densities associated with general quantum
mechanical states, the answer is \textit{bounded} ones,
denoted ${\mathcal L}^\infty(\MM)$, because otherwise we are not assured that the
integral in (\ref{eq:integral-f-Gamma}) exists.
Thus, we pass to a second stage with
\begin{equation}
  \label{eq:Obs_1}
\Obs_1 \defeq {\mathcal L}^\infty(\MM).
\end{equation}
Relative to this class of observables, the state simply \textit{is} a density,
specifically, $\dens\Gamma$ in (\ref{eq:integral-f-Gamma}).
The states are now
\begin{equation}
  \nonumber
\Sts_1 \defeq {\mathcal L}^1(\MM)_{+,\NN},
\end{equation}
non-negative integrable functions with total integral $\NN$,
and the observable-state pairing is
\begin{equation}
  \label{eq:DFT-pairing}
\pair{f}{\rho} = \int_\MM f(x) \rho(x)\, dx,  
\end{equation}
which satisfies
\begin{equation}
  \nonumber
\pair{f}{\rho} = \pair{\Num(f)}{\Gamma},
\end{equation}
whenever $\Gamma \in {\dens}^{-1}\rho$.
The pairings on the LHS and RHS are not literally the same thing, but
are realizations of the same abstract idea in two different settings.

For DFT, we want to modify this structure somewhat, by restricting to
densities coming from states of finite kinetic energy, and considering
nonlinear observables.

\subsubsection{finite kinetic energy}

In the general QM context, a single-particle wavefunction which is, for example,
a nonzero constant over a cubical region and zero outside, is legitimate,
but it has infinite kinetic energy. (Physically, this is pretty clear.
Mathematically, we extend the expectation of the kinetic energy operator outside
its ordinary domain by saying it is $+\infty$ there. This is unambiguous because
kinetic energy is bounded below.)
However, we want to insist on finite kinetic energy, and
this entails a state space smaller than $\Sts_1$.
We will denote this set of densities by $\SD$;
Lieb\cite{Lieb83} calls it $\mathscr{I}_\NN$.
Precisely, the additional requirement for $\rho$ to be in $\SD$ is that $\nabla \sqrt{\rho}$
be square integrable, so
\begin{equation}
\label{eq:Sts-KE}
\Sts_2 \defeq \setof{\rho\in \Sts_1}{\nabla\sqrt{\rho}\;\text{ is square-integrable}}.
\end{equation}
Correspondingly, the space of observables can be expanded.
In fact, the integral $\int v\rho\, dx$ is well-defined for every $\rho\in\SD$
not only when $v$ is essentially bounded, but also when $|v|^{3/2}$ is integrable.
Thus, 
\begin{equation}
\label{eq:Obs-KE}
\Obs_2 \defeq {\mathcal L}^\infty(\MM) + {\mathcal L}^{3/2}(\MM). 
\end{equation}

\subsubsection{Nonlinear observables and intrinsic energy}\label{sec:intrinsic-energy}

Now suppose $A$ is any bounded operator.
We can certainly associate the set $\setof{\Tr \Gamma A}{\dens\Gamma = \rho}$ with $\rho$.
In case $A$ represents an energy, it is physically well-motivated to associate
the infimum of this set to $\rho$. That works even if $A$ is only
bounded below, like kinetic energy.
Define, therefore,
\begin{equation}
\label{eq:F0(rho)}
F_0(\rho) \defeq \inf \setof{ \Tr \hat{T}\Gamma }{\dens\Gamma =\rho}.
\end{equation}
This makes sense for all densities. For some $\rho$, \hbox{$F_0(\rho) = +\infty$}
by this definition. Those densities for which $F_0$ is less than $+\infty$
is called the \textit{effective domain}, denoted $\dom F_0$.
It is exactly $\SD$.
Because $\dens$ is a linear map, it follows from (\ref{eq:F0(rho)}) that
$F_0$ is convex:
\begin{align}
  0\le s & \le 1 \Rightarrow
                 \nonumber \\
& F_0((1-s)\rho+s\rho') \le (1-s) F_0(\rho) + s F_0(\rho').
\end{align}
Because $F_0$ is bounded below, it does not matter whether $\rho$
and $\rho'$ are in $\dom F_0$ ($\infty + a = \infty$ if $-\infty < a$).

$F_0$ is the noninteracting \textit{intrinsic energy} (functional).
If $\hat{W}$ is an interaction between the particles, then we can
analogously define an interacting intrinsic energy ($F$, no subscript) with
$\hat{T}$ in (\ref{eq:F0(rho)}) replaced by $\hat{T} + \hat{W}$.
Assuming $\hat{W}$ is relatively bounded with respect to $\hat{T}$,
e.g., Coulomb interaction, $\dom F = \dom F_0$.

\subsubsection{Constrained search and Legendre-Fenchel transform}\label{sec:constrained-search}

Now, suppose ${v}$ is some external one-body potential.
The minimum energy of states with density $\rho$ in presence of $v$ is
$F(\rho) + \int {v}\rho\, dx$. Thus, if there is a ground state, the
ground state energy is
\begin{equation}
\label{eq:E-min-version}
E({v}) = \min
\setof{ F(\rho) + \int {v}\rho\, dx }{\rho\in\SD}
\end{equation}
This embodies the central, appealing, idea of the constrained-search
formulation\cite{Levy79,Lieb83}. 
The minimum will not exist, and $E(v)$ will not be defined, if there are no ground states.
This is not an exotic possibility; it occurs for a constant potential on $\Real^3$.
That problem is easily fixed by replacing $\min$ by $\inf$.
Even so, what is the domain of $E$?
Consider that the integral $\int v\rho \, dx$ is well-defined, for a trap potential, i.e.,
bounded below and satisfying $v(x) \to \infty$ as $|x| \to \infty$.
For some densities, the integral has the value $+\infty$, for others it is finite.
We will rule these out, however, requiring the integral to be finite for every $\rho$.
One might regard this a valid physical requirement as it stands.
Another reason, discussed below, is that potentials play the role of derivatives
of $F$. They should therefore be unambiguously integrable against differences of
densities.
Thus, we arrive at the conclusion that the sensible space of potentials is
precisely $\Obs_2$ (\ref{eq:Obs-KE}), times a unit of energy.
For notational simplicity (and forgetting about the energy unit), we give this space a new name:
\begin{equation}
  \label{eq:SV}
\SV \defeq {\mathcal L}^\infty(\MM) + {\mathcal L}^{3/2}(\MM).
\end{equation}
In a more abstract context, we continue to use the symbol $\SV$ to represent whatever
vector space plays this role.

The integral in (\ref{eq:E-min-version}) is thus our previously introduced pairing,
giving us the final form
\begin{defn}[ground energy]\label{def:E}
The \textit{ground energy} of $v\in\SV$ is 
\begin{equation}
\label{eq:E from F}
E(v) = \inf \setof{ F(\rho) + \pair{v}{\rho} }{\rho\in\SD},  
\end{equation}
\end{defn}
As the infimum of a collection of linear functionals, the ground energy is
automatically \textit{concave}, i.e., $-E$ is convex.
For a concave functional, the effective domain is defined oppositely from that
for a convex functional (i.e., where it is greater than $-\infty$).
Although not obvious on its face, $E(v) > -\infty$ for \textit{every}
$v\in\SV$. So, $\dom E = \SV$.

Now, in case there is a ground state for $v$, a basic idea of calculus suggests
that the minimum of the RHS of (\ref{eq:E from F}) should have a differential
characterization, e.g.
\begin{equation}
\nonumber  
\diff F(\rho) + {v} \stackrel{?}{=} 0
\end{equation}
for some kind of derivative $\diff$.
Therefore, we turn next to the problem of differentiation of functions
in the context of a dual pair of vector spaces.
Although $\SD$ is not a vector space, if we are interested not only in
densities, but arbitrary multiples of differences of densities, the
vector space generated by $\SD$, denoted $\vecsp\SD$ comes in naturally.
Then, we also need to extend the intrinsic energy to $\vecsp\SD$.
Some elements of $\vecsp\SD$ which are not in $\SD$ are densities for which
$F$ and $F_0$ are already defined as $+\infty$. Now we give that value to all
the others, as well. This maintains convexity of those functionals.

\section{Interlude: interpretations}\label{sec:interpretations}

The previous section sketched a view of standard DFT, and introduced most of the
main characters: $F_0$, $E_0$, $F$, and $E$. Others derived from these, such as
the Hartree-exchange-correlation energy $\Phi$, will be added later.
These are all unambiguously, though not accessibly, defined in terms of the
quantum mechanics of many body systems in Euclidean space. They are, in the jargon,
\textit{exact}.

However, approximations, or perhaps better to say, \textit{models}, are inevitable
in this business, and we would like to draw conclusions applicable to computations
with these models which do not depend in some uncontrolled way on
their being close enough to exact, in some vague sense.
This motivates an axiomatic approach. We identify key properties of the
standard (``exact'') interpretation and proceed assuming only that our functionals,
and the spaces they are defined on, have those properties.
The list of assumptions (axioms/postulates) will grow in a couple of stages, so
that later sections assume more.
In addition to approximate exchange-correlation energy functionals, this makes
room for other kinds of deviations from the standard interpretation, such
as a system living on a torus rather than Euclidean space, a background
confining potential, nonzero temperature, extra degrees of freedom
(e.g., spin DFT, current DFT).
A concrete system satisfying the axioms is referred to as an \textit{interpretation}.
So, for example, the exact noninteracting functional $F_0$ with
local density approximation for exchange-correlation energy defines a
interpretation.

\section{A unilateral derivative}\label{sec:derivatives}

Starting in this section, we begin to work in a relatively abstract way.
For instance, instead of the specific vector spaces $\vecsp \SD$ and
$\SV$ in (\ref{eq:SV}),
we say simply that we have a pair of vector spaces
$\SV$ and $\SX$ with a nondegenerate pairing $\pair{\cdot}{\cdot}$
(See Def.~\ref{def:dual-system}).

\subsection{Motivation}

We pursue here the idea that the essence of \textit{derivative} is some sort of
(local) linear approximation, what kind of approximation being open to discussion.
For example, the derivative of a smooth function $\Arr{\Real^2}{f}{\Real}$ 
is packaged as a linear functional through its gradient.
The graph of the affine function
\hbox{$x' \mapsto f(x) + \nabla f(x) \cdot (x'-x)$}
is tangent to the graph of $f$ at $x$, and in that sense consitutes the
best affine approximation to $f$ near $x$.
The dot product is a pairing of ${\Real^2}$ with itself,
$\pair{x}{y} = x\cdot y$, so we might also write this as
\hbox{$f(x) + \pair{\nabla f(x)}{x'-x}$}.

Now, suppose that $f$ is not smooth, for example, $f(x) = |x|$, at the origin.
If our interest is in minimization, a \textit{one-sided} kind of approximation
can be perfectly suitable. If $|n| \le 1$, the graph of the linear functional
\hbox{$x \mapsto \pair{x}{n}$} touches that of $f$ at $x=0$ and is nowhere above
it. In that weak sense, it is a kind of linear approximation.
Because there is not just a single $n$ which works here, we see that when we
relax our notion of approximation in this way, we can end up with derivatives
which are \textit{set-valued}.

\subsection{Lower and upper semiderivatives}\label{sec:lower-upper}

We will define derivatives for dual systems.
\begin{defn}[dual system]\label{def:dual-system}
A \textit{dual system} consists of a pair
of vector spaces $\SV$ and $\SX$ and a map
\hbox{$\Arr{\SV\times\SX}{\pair{\cdot}{\cdot}}{\Real}$}
which is linear in each variable with the other held fixed, and such that
for every $x\in\SX$, there is $v\in\SV$ (and for every $v\in\SV$ there
is $x\in\SX$) such that $\pair{v}{x} \neq 0$.

For a compact notation, we denote this dual system by $\pair{\SV}{\SX}$
\end{defn}
Nondegeneracy is the new concept in this definition.
Essentially it means neither space involved is ``too small'', since
it says that $x\in\SX$ can be unambiguously identified by
the values of $\pair{v}{x}$ as $v$ ranges over $\SV$, and vice versa.

Now we can define a unilateral notion of derivative relative to a dual system.
Putting a bar above or below `$\Real$' indicates augmentation by
$+\infty$ or $-\infty$, for instance,
$\overline{\underline\Real} = \Real\cup\{-\infty,+\infty\}$,
and $\varliminf$ denotes limit inferior ($\liminf$).
\begin{defn}\label{def:semiderivative}
  The \textit{lower semiderivative} of $\Arr{\SX}{f}{\overline{\underline\Real}}$ at $x$
[with respect to the pairing 
\hbox{$\Arr{\SV\times\SX}{\pair{\cdot}{\cdot}}{\Real}$}]
  is the \textit{set} of $v \in\SV$ such that
\begin{equation}
\label{eq:asymptotic-subgrad}
\varliminf_{s \downarrow 0} \frac{f(x+su) - f(x)}{s} \ge \pair{v}{u},
\text{ for all } u\in \SX.
\end{equation}
The lower semiderivative is denoted $\ldiff f(x)$.

The \textit{upper semiderivative}, $\udiff{f}(x)$ is defined
by an analogous equation with $\varliminf$ replaced by $\varlimsup$,
and $\ge$ by $\le$.

Similarly, exchanging the roles of $\SX$ and $\SV$, we obtain
semiderivatives of functions on $\SV$ with respect to the same pairing.
\end{defn}

\subsection{Remarks}\label{sec:derivative-remarks}

\begin{enumerate}
\item
 Geometrically, $u \in \ldiff f(x)$ means that the
hyperplane $y \mapsto f(x) + \pair{u}{y}$ in $\SX\times \Real\cup{\pm\infty}$
is asymptotically not above the graph of $f$ as $y\to x$.

\item 
If $v$ is in both $\subdiff f(x)$ and $\supdiff f(x)$, then
$\lim_{s \to 0} s^{-1}[{f(x+su) - f(x)}] = \pair{v}{u}$.

\item
For a \textit{convex} function $\Arr{\SX}{f}{\overline\Real}$,
$\subdiff f$ has a much simpler characterization, and it
does not involving limits at all.
$v\in\subdiff f(x)$ precisely when, for all $y$,
\begin{equation}
  \nonumber
f(x) + \pair{v}{x} \le f(y) + \pair{v}{y}.  
\end{equation}
For application to DFT, this would suffice for $F_0$, $F$, $E_0$ and $E$.
However, in Kohn-Sham approach, we deal with 
$\Phi$. defined by $F=F_0 + \Phi$, and this cannot be assumed to be
either convex or concave.

\item
  If $u$ is in $\ldiff f(x)$ and $v$ is in $\ldiff g(x)$, then
  \hbox{$u+v \in \ldiff (f+g)(x)$}. This follows since the limit
  inferior of a sum is at least as large as the sum of the limits inferior,
  but the geometric description of the first item might be an easier way.
  Beware! This does not work with subtraction, that is,
  $u \in \ldiff f(x)$ and \hbox{$u+v \in \ldiff (f+g)(x)$} do \emph{not} imply 
  $v\in \ldiff g(x)$.

\end{enumerate}

\section{Kohn-Sham machines}\label{sec:KS-machines}

The top level of a Kohn-Sham computation involves densities and potentials
alone, with no explicit reference to quantum mechanics.
This Section abstracts that top level perspective as a \textit{Kohn-Sham machine},
offering a limited menu of operations on potentials and densities,
and provided by modules which are regarded as black boxes.
The following Section then analyzes the question, 
given an external potential $\tgt{v}$, how can those operations be harnessed
to make progress toward finding an interacting ground density for
$\tgt{v}$? This will be given an abstract phrasing, and we will have to
find an appropriate sense of \textit{progress} to deal with it.

\subsection{Postulates}\label{sec:KS-assumptions}

We abstract the situation described in the preceding section
in the form of the following assumptions.

\begin{enumerate}[label=A{\arabic*}.,series=aaxs,ref=A\arabic*]
\item \label{A:D}
$\SD\subset \vecsp\SD$ is convex
\item \label{A:cvx}
$\Arr{\SD}{F}{\Real}$ is a convex function, bounded below
\item \label{A:dual-pair}
{$\Arr{\SV\times\vecsp\SD}{\pair{}{}}{\Real}$}
is a nondegenerate bilinear pairing
of a second real vector space $\SV$ with $\vecsp\SD$.
\end{enumerate}
These are just the beginning.
Additional axioms refining the set-up will be added in Sections \ref{sec:RCI}
and \ref{sec:RCII} as their desirability becomes clear. In all cases, they 
reflect properties of exact functionals in the standard interpretation.
Postulates
\ref{A:D}, \ref{A:cvx}, and
\ref{A:dual-pair} are
\emph{descriptive}. 
In this section and the next, we assume that there is a second function
$F_0$ satisfying \ref{A:cvx}.
Moreover, we will have \emph{computational/procedural} assumptions on
$F_0$ and $\Phi \defeq F - F_0$ as specified in section \ref{sec:feasibility}.
Those will have no direct relevance for the development following section
\ref{sec:feasibility}.

The functions $F_0$, $F$, and $\Phi$ are extended to all of
$\vecsp \SD$ by setting them equal to $+\infty$ off of $\SD$.
This is a matter of \textit{convention}, designed to maintain convexity of
$F$ and $F_0$ and the equality
\begin{equation}
  \label{eq:F=F0+Phi}
F = F_0 + \Phi,
\end{equation}
while creating no barriers to lower semi-differentiability.

\subsection{Standard interpretation}

The standard interpretation is that $\SD$ is the set of densities of finite intrinsic
energy introduced in Section \ref{sec:to-DFT}, i.e., $\SD = \dom F_0 = \dom F$,
\hbox{$\SV = \mathcal{L}^\infty(\mathcal{M}) + \mathcal{L}^{3/2}(\mathcal{M})$},
$F_0$ is the noninteracting (and $F$ the interacting) intrinsic energy.
That \ref{A:D} -- \ref{A:dual-pair} are satisfied in the standard interpretation
was already established in Section \ref{sec:to-DFT}.
Ground energy $E$ is defined from and $F$ by (\ref{eq:E from F}).
In these abstract terms, the Standard Problem of finding a ground density for potential
$\tgt{v}$ can be phrased as: find \hbox{$\rho \in \supdiff E(\tgt{v})$}.
We will find a different formulation more useful and enlightening.

However, whereas what follows Section \ref{sec:progress} has some interest
in the case of the standard interpretation, where everything is exact,
that is hardly true in this Section, and the next.
These two Sections are so closely tied to the mechanics and possibilities
of Kohn-Sham computation that the realistic attitude with which to read them is that
$F_0$ is exact, while $\Phi$ is a model Hartree-exchange-correlation energy,
constrained only by the requirement that $F$ satisfy \ref{A:cvx}.
The term \textit{Hartree-exchange-correlation} energy indicates that $\Phi$
is considered physically (interpreted) to consist of three parts: the classical Coulomb interaction
energy of the given charge density (Hartree), and two quantum effects, exchange and
correlation. Often $\Phi$ is split explicitly into the Hartree energy, which is simple,
unambiguous and explicit, and the exchange-correlation (XC) energy, the part which is really
approximated. For our purposes, however, it makes more sense to take it as a unit. 

\subsection{Excess energy}\label{sec:excess-energy}

Intrinsic energy $F$ is a function of density, ground energy $E$ of potential.
Underlying our approach is the idea that it is fruitful to think in terms of
both density and potential simultaneously.
This means that we mostly think of things as functions on the product
space $\SV \times \SD$, and package $F$ and $E$ together into the
{\it excess energy}
\begin{equation}
\label{eq:excess-energy}
\xs(v,\rho) \defeq F(\rho) + \pair{v}{\rho} - E(v) \ge 0.
\end{equation}
$\xs(v,\rho)$ answers the question,
``how close to the ground energy $E(v)$ can one get with states of density $\rho$?''
and is convex in each variable, holding the other fixed.
The zero set
$\zero \defeq \setof{(v,\rho)\in\SV\times\SD}{\xs(v,\rho) = 0}$
(simply $\{\xs=0\}$ in abbreviated form) contains all possible solutions
of all possible ground density problems. If $(v,\rho)$ is in $\zero$,
we call it a {\it ground pair}.

\begin{figure}
  \centering
\includegraphics[width=80mm]{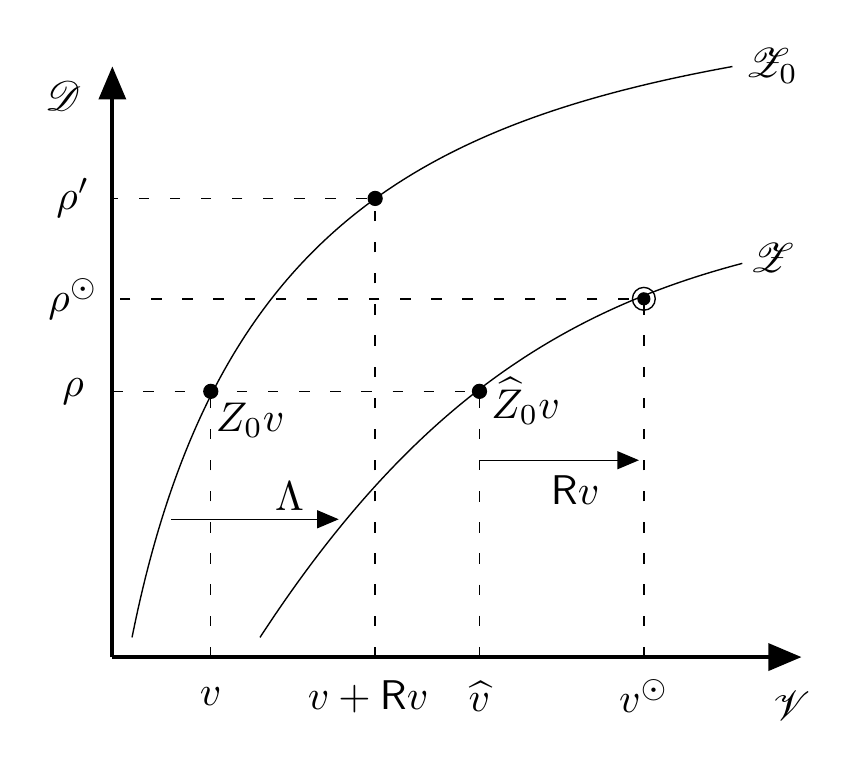}
\caption{
  Schematic representation of the bivariate perspective in the product space $\SV\times\SD$.
  The zero excess energy sets $\zero_0$ and $\zero$ are indicated, along with some of the
  functions listed in Table \ref{tab:feasible}.
  Of course, the picture is unfaithful in some aspects:
  $\SV$ and $\SD$ are generally infinite-dimensional,
  $\zero$ and $\zero_0$ are not likely to be smooth, or even
  (single-valued) functions.
\label{fig:bivariate}}
\end{figure}

Noninteracting versions, $E_0$, $\xs_0$, and $\zero_0$ are defined
from $F_0$ in the same way as $E$, $\xs$ and $\zero$ from $F$.
In distinguishing between the two, we prefer the more neutral designations
\textit{reference/perturbed} to \textit{noninteracting/interacting}.
Fig.~\ref{fig:bivariate} depicts, in a cartoon way, the zero sets
in the product space $\SV\times\SD$.

\subsection{Primitive operations and feasibility}\label{sec:feasibility}

Some of the functions of the theory listed above, e.g., $F$,
are not provided in modular form by ordinary DFT software.
This is the reason why it is an interesting to ask about strategies to solve
the basic problem.
The menu of primitive operations consists of: solution of the noninteracting
problem, computation of HXC energy and potential, and calculation of the
integral $\int v(x) \rho(x) \, dx$. In our more neutral language, they are
given in Table~\ref{tab:prim-ops}.
\begin{table}[h]
  \centering
  \begin{tabular}{lr}
    operation & standard interpretation \\
    \hline
${E_0}$ & noninteracting ground energy \\
$\slct{\supdiff E_0}$ & (one) noninteracting ground density \\ 
$\Phi$ & HXC energy \\
$\slct{\subdiff\Phi}$ & (one) HXC potential \\
$\pair{\cdot}{\cdot}$ & potential-density pairing
\end{tabular}
\caption{Primitive feasible operations of the Kohn-Sham machine.
Lower and upper semiderivatives are set-valued.
The notation $\slct{\cdot}$ means that we can find out if the set is
nonempty, and get (at least) one member if so.
}\label{tab:prim-ops}
\end{table}
The primitive operations, as well as anything achievable by a finite combination of
them, is \textit{feasible}.
We will mostly be engaged in demonstrating feasibility by exhibiting appropriate such
combinations. Section \ref{sec:infeasible-strategy} makes a soft claim of infeasibility,
but it must be recognized that such claims are significantly trickier, and 
potentially subject to criticism on the grounds that our list of primitive operations
is incomplete. 
Certainly, nothing here should be construed as making claims about what completely
different methods, such as quantum Monte Carlo, can do.

\subsection{Generating ground pairs}\label{sec:ground-pairs}

From the primitive operations (Table~\ref{tab:prim-ops}), we will now synthesize some
new feasible operations which allow generation of reference and perturbed
ground pairs, and which may be useful in solving the Standard Problem.
They are listed in Table~\ref{tab:feasible} and
some are illustrated in a schematic way on Fig.~\ref{fig:bivariate}.
\begin{table}[h]
  \caption{Basic feasible functions/operations, described in the text.
    $\circ$ is the composition operator, $\pi_{\SV}$ extracts $\SV$ component,
    and $\rightharpoonup$ indicates a partial (not everywhere defined) function.
    \label{tab:feasible}
    }
  \centering
\begin{tabular}{lcl}
  \hline
  name & definition & type \\ \hline
  $Z_0$  & $v \mapsto (v, \slct{\udiff E_0(v)} )$ &
 $\SV \rightharpoonup \zero_0$ 
 \\  
   $\KS$ & $(v,\rho) \mapsto (v-\slct{\ldiff\Phi(\rho)}, \rho)$ &
 $\zero_0 \rightharpoonup \zero$
   \\
 $\wh{Z}_0$ & $\KS \circ Z_0$ &
 $\SV \rightharpoonup \zero$
    \\
  $(\wh{\phantom{v}})$
  & $\pi_{\SV}\circ \wh{Z}_0 $ &
 $ \SV \rightharpoonup \SV$
  \\
$\Rsd_{\tgt{v}}$  
& $v \mapsto \tgt{v} - \wh{{v}}$  &
 $ \SV \rightharpoonup \SV$
   \\
  $F^\tinyHK$
  & $(v,\rho) \mapsto E_0(v)-\pair{v}{\rho}+\Phi(\rho)$
  & $\zero_0 \rightarrow \Real$
   \\
  $E^\tinyHK$
  & $(v,\rho) \mapsto F^{\tinyHK}(v,\rho) + \pair{\wh{v}}{\rho}$
  & $\zero_0 \rightharpoonup \Real$
\\ \hline
\end{tabular}
\end{table}

Let us consider these operations.
$Z_0$ is a trivial rephrasing of $\slct{\udiff E_0}$;
it merely pairs a potential with a corresponding reference system
ground density. It is not a map into densities $\SD$, but
into the subset $\zero_0$ of the product space $\SV\times\SD$.
$\KS$ puts $\ldiff\Phi$ to work, and is more interesting.
\hbox{$(v,\rho)\in\zero_0$} is equivalent to
\hbox{$F_0(\rho) + \pair{v}{\rho} \le F_0(\rho') + \pair{v}{\rho'}$},
which is equivalent
to \hbox{$-v\in \ldiff F_0(\rho)$}, by remark 3 of Section \ref{sec:derivative-remarks}.
An analogous statement holds for $F$ and $\zero$.
Since $F = F_0 + \Phi$, remark 4 of Section \ref{sec:derivative-remarks} gives
\begin{equation}
(v,\rho)\in\zero_0 \Rightarrow (v-\slct{\ldiff\Phi(\rho)},\rho)\in\zero.
\nonumber
\end{equation}
The reverse implication is not valid.
Summing up: given $v\in\SV$, $Z_0 v$ is a reference ground pair, and
\hbox{$\wh{Z}_0 v = \KS \circ Z_0 \, v = \KS (Z_0 \, v)$}, when it exists,
is a perturbed ground pair with $\SV$ component \hbox{$\wh{v}$}. That is
if \hbox{$(v,\rho)\in\zero_0$} then \hbox{$(\wh{v},\rho)\in\zero$}, as shown
in Fig.~\ref{fig:bivariate}.
To see the point of this, recall that
our Standard Problem is to find a point on $\zero$ with specified first component
$\tgt{v}$, so we are naturally interested in how close $\wh{v}$ is to $\tgt{v}$.
The map $\Rsd_{\tgt{v}}$ supplies that information. Usually, we will suppress
$\tgt{v}$ for notational simplicity.

These functions are all \textit{partial}, which is why the Table contains
`$\rightharpoonup$' rather than `$\rightarrow$' in the type column.
Certainly, some potentials have no ground density. For example, the uniformly
zero potential in $\Real^3$. Given that partiality, there is no benefit for us to
assuming that $\slct{\subdiff \Phi}$ is total.
Computationally, our assumption is that an exception, rather than garbage, is
returned in case there is no value.

The perspective revealed here is different from the usual one.
Ground pairs are the only points in $\SV\times\SD$ which are usefully
accessible. Reference ground pairs can be feasibly selected by their first component,
but perturbed ground pairs only in a distorted kind of way.
The common talk of ``self-consistency'' seems inappropriate from this
perspective. Points on $\zero$ generated by using the basic operations
are certainly not \textit{inconsistent} in any sense. Their only possible
defect is not being one that we want.
The question then, is how to use the
expanded stock of basic operations in Table~\ref{tab:feasible} to find a
suitable pair, that is, one solving the Standard Problem.
The next section takes up the question of how to make progress toward that goal.
First, we discuss the last row of the table.

\subsection{HK maps}\label{sec:HK-maps}

The Table \ref{tab:feasible} entries discussed to this point use only
$\slct{\udiff{E_0}}$ and $\slct{\ldiff{\Phi}}$ from the primitives of
Table~\ref{tab:prim-ops}.
The others (namely, $\Phi$, $E_0$ and $\pair{\cdot}{\cdot}$)
are needed for $F^{\tinyHK}$ and $E^\tinyHK$.
If $(v,\rho) \in \zero_0$, then
\hbox{$0 = \xs_0(v,\rho) = F_0(\rho)+\pair{v}{\rho}-E_0(v)$}, and therefore
\begin{equation}
\nonumber
F(\rho) = F^{\tinyHK}(v,\rho) \defeq E_0(v)-\pair{v}{\rho}+\Phi(\rho).
\end{equation}
The superscript {\tiny{HK}}, standing for `Hohenberg-Kohn',
is there because this is much closer to the original\cite{Hohenberg+Kohn} intrinsic energy
(``universal functional'') definition of Hohenberg and Kohn than the later
constrained-search formulation\cite{Levy79,Levy82}.
The point is that auxiliary data consisting of a potential partner in the
{reference} system is needed to obtain $F(\rho)$.
Since $(\wh{v},\rho)$ is a perturbed ground pair, once we have $F(\rho)$,
the interacting ground energy of $\wh{v}$ is obtainable as
\begin{equation}
\nonumber
E(\wh{v}) = E^{\tinyHK}(v,\rho) \defeq F^{\tinyHK}(v,\rho)+ \pair{v}{\rho}.
\end{equation}

\subsection{Reduced KS-machine}\label{sec:reduced-KS-machine}

Generally, the term \textit{Kohn-Sham machine} refers to any collection of feasible
operations, such as those in Table~\ref{tab:feasible}.
It is easier to focus on the essentials, though, if we consider
a \textit{reduced KS-machine} offering the single operation
\begin{align}
  \text{input:}\quad & v \nonumber \\
  \text{output:}\quad & (\wh{v},\rho) \in \zero, \, E(\wh{v}), \, F(\rho)
                        \label{eq:reduced-KS}
\end{align}
This is straightforwardly constructed from those in Table~\ref{tab:feasible}.
$E(\wh{v})$ and $F(\rho)$ come from the HK maps.

One use of the reduced KS-machine gives us a perturbed ground pair in $\zero$,
and its essential characteristics. 
The only problem is that it is unclear how to control \emph{either} its
potential or its density component.

\section{Verifiable progress}\label{sec:progress}

Essentially, the only feasible access to $\zero$ is via $\wh{Z}_0$.
The picture of the previous section suggests the following approach to
the Standard Problem.
Pick a potential $v$ (somehow), obtain $\wh{Z}_0 v$, compare its first
component to $\tgt{v}$, if the difference $\Rsd v$ is not satisfactorily
small, choose a new input to $\wh{Z}_0$ based on the experience.
Repeat until satisfied.
This section is concerned with how to make that choice of next input so
that some form of progress is assured. 

\subsection{Progress}

Suppose we generate a sequence of points $(v_n,\rho_n)$ on $\zero$.
How would we ascertain that we were making progress toward the solution
to the Standard Problem? 
One interpretation would be that $v_n$ and $\rho_n$ are converging to the
target potential and density. However, the latter is unknown.
We could ask if $\tgt{v} - v_n$ is becoming small, but 
that requires a quantitative measure of the ``size'' of a potential difference.
We defer such topological considerations to the following sections, in order to see
what can be done without them. 

Fortunately, the basic feasible operations in hand already provide
the means to assess whether one density is \textit{energetically} better
than another, provided we have them in the form of components of points
on $\zero_0$ or $\zero$.
The energetic measure of how close $\rho$ is to a ground density for
$\tgt{v}$ is $\xs(\tgt{v},\rho)$.
So, define
\begin{align}
  \nonumber
  \inc(\tgt{v};\rho,\rho') & \defeq \xs(\tgt{v},\rho) - \xs(\tgt{v},\rho')
                             \nonumber \\
 &\;\; = F(\rho)-F(\rho')+\pair{\tgt{v}}{\rho-\rho'}.
\end{align}
If this is less than zero, $\rho$ is a ``better'' density than $\rho'$,
indicating that going from $\rho'$ to $\rho$ is \textit{progress} of a sort.
The important point is that
\begin{align}
(v,\rho),(v',\rho')\in\zero_0 & \Rightarrow
\\ 
 \inc(\tgt{v};\rho,\rho') =
 & F^{\tinyHK}(v,\rho)-F^{\tinyHK}(v',\rho')+\pair{\tgt{v}}{\rho-\rho'}.
    \nonumber
\end{align}
Evidently, this is feasible. It is the measure of progress we will use in this section. 

\subsection{Conventional fixed-point formulation}\label{sec:usual-strategy}

Given $v_0$ as input, the KS-machine produces (barring exceptions)
a reference ground pair $(v_0,\rho_0)= Z_0 v_0$
and a perturbed ground pair $(\wh{v_0},\rho_0)= \wh{Z}_0 v_0$ as output.
For purposes of comparing with the usual formulation of KS iteration,
it may be helpful to refer to $\rho_0$ and $\wh{v_0}$ as the 
\textit{output density} and \textit{output potential}, respectively.
Now, in that situation, a simple idea for the next input is
\begin{equation}
  \label{eq:strategy-0}
v_1 = v_0 + \Rsd v_0.  
\end{equation}
The pattern can be continued to entire sequence
\hbox{$(v_n,\rho_n,\wh{v_n})_n$}, with
\hbox{$v_n + \Rsd v_n = v_{n+1}$},
\hbox{$\wh{v_n} + \Rsd v_n = \tgt{v}$}.
Unpacking definitions shows that this is equivalent to
\begin{equation}
v_{n+1} = \tgt{v} + \slct{\subdiff\Phi(\rho_n)}, 
\end{equation}
and (\ref{eq:strategy-0}) is thereby revealed to be the usual naive iteration step.
This is labelled ``naive'' because it is a well-known empirical fact
that this scheme is subject to problems which can be ameliorated by \textit{mixing}.
``Charge-sloshing'', for instance, is a situation where, from iteration to iteration,
the density gets stuck alternating between two fairly well-defined but distinct densities.
In the bivariate perspective being built here, mixing in general would be expressed as the idea
that (\ref{eq:strategy-0}) is a good ``direction'' in which to shift the
input potential, but that maybe a more cautious step is advisable:
\begin{equation}
  \label{eq:strategy-mixing}
v_1 = v_0 + \lambda \Rsd v_0, \quad 0 < \lambda \le 1.  
\end{equation}
Conventionally, the same rough idea is implemented differently.
An auxiliary ingredient, an \textit{input density} is introduced to
parametrize the input potential, as
\begin{equation}
  v_{n+1} = \tgt{v} + \slct{\subdiff\Phi(\rho^{\text{in}}_{n+1})},
\end{equation}
and mixing is done on the auxiliary quantity:
\begin{equation}
  \label{eq:strategy-density-mixing}
 \rho^{\text{in}}_{n+1}
= \lambda \rho_{n} +(1-\lambda) \rho^{\text{in}}_{n}.
\end{equation}
This kind of parameterization gives rise to the apparently common view that
Kohn-Sham theory \textit{intrinsically} involves a fixed-point problem,
i.e., of the map $\rho_n^{\text{in}} \mapsto \rho_n$.
From the bivariate perspective, that is entirely incidental.
It is unclear what advantages it may have over working directly with potentials
as in (\ref{eq:strategy-mixing}).
Most importantly for the present work, I am unable to prove anything about such
schemes, whereas favorable results will be obtained for something like (\ref{eq:strategy-mixing}).

\subsection{Utilities}\label{sec:utilities}

We collect some useful identities, proven by straighforward
manipulation, which will be used in Sections
\ref{sec:infeasible-strategy} and \ref{sec:feasible-strategy}.
Items {1} -- {3} hold for either the reference system
(in which case subscripts $0$ should be attached) or the perturbed system.
They are entirely elementary and depend only on convexity properties of $F$ and $E$.
Recall the definition of excess energy:
\begin{equation}
  \nonumber
\xs({v},\rho) = F(\rho) + \pair{{v}}{\rho} - E({v}).
\end{equation}

\smallskip
\noindent {1}.
{Cross-difference identity:}
\begin{align}
\xs(v,\rho)  
+ \xs(v',\rho')   
=& \xs(v,\rho')  
+ \xs(v',\rho)  
\nonumber \\
& + \pair{v - v'}{\rho - \rho'}. 
\label{eq:cross-difference}
\end{align}
Each of $v$, $v'$, $\rho$, and $\rho'$ appears 
once in a $\xs$ on either side. 
Upon substituting the definition of $\xs$,
all $F$'s and $E$'s cancel out, leaving only
potential-density pairings.

\smallskip
\noindent {2.} Monotonicity:
\begin{equation}
(v,\rho), (v',\rho')  \in \zero \; \Rightarrow \;
 \pair{v - v'}{\rho - \rho'} \le 0.
\label{eq:monotonicity}
\end{equation}
If either $(v',\rho)$ or 
$(v,\rho')$ fails to be a ground pair, then the inequality is strict.
This monotonicity inequality\cite{Phelps88,Aubin+Ekeland,Laestadius+18-JCP}
is an immediate specialization of the cross-difference identity (\ref{eq:cross-difference}).
It generalizes an inequality previously derived in a specificly
DFT context\cite{Gritsenko+Baerends-04,Wagner+13}.

\smallskip
\noindent {3.}
\begin{align}
  (v,\rho)\in\zero & \; \Rightarrow
                     \nonumber \\
    & \xs({v}',\rho) = E({v}) - E({v}')  + \pair{{v}'-{v}}{\rho}.
\label{eq:v-shift}
\end{align}
Expand $\xs({v}',\rho) -  \xs({v},\rho)$ using the definition of $\xs$.

\smallskip
\noindent {4.}
Assuming $\rho\in\dom \ldiff\Phi$,
and with $\ldiff_\rho$ denoting the semiderivative with respect to $\rho$ at fixed $v$,
 \begin{equation}
  (v,\rho)\in\zero_0  \;\Rightarrow\;
 \Rsd v \in \ldiff_\rho\xs(\tgt{v},\rho).
\label{eq:Rv-in-subdiff}
\end{equation}
According to the definition of excess energy,
$\ldiff_\rho \xs(\tgt{v},\rho) = \ldiff F(\rho) + \tgt{v}$.
Since $(v,\rho)\in\zero_0$ implies that $-\wh{v} = \Rsd v - \tgt{v} \in \ldiff F(\rho)$,
the conclusion follows.

\subsection{An infeasible strategy}\label{sec:infeasible-strategy}

Given $v_0$, define $v_1$ as in (\ref{eq:strategy-0}), i.e., $v_1 = v_0 + \Rsd v_0$.
Corresponding densities are defined by the conditions
\begin{equation}
  \label{eq:(v,rho)-ends}
(v_0,\rho_0),\, (v_1,\rho_1) \in \zero_0.  
\end{equation}
Now, we consider two ideas for interpolation.
The first is defined by a linear interpolation in density.
Assuming $\rho_1 \neq \rho_0$, it makes sense to define
\begin{equation}
  \label{eq:rho-interp}
{\rho}_\lambda =  (1-\lambda){\rho}_0 + \lambda {\rho}_1,
\end{equation}
for $0\le\lambda$.
A form of the following Proposition
appears to have been first given by Wagner \hbox{\textit{et al.}}\cite{Wagner+13},
later corrected and rigorized by Laestadius \hbox{\textit{et al.}}\cite{Laestadius+18-JCP}.
At first sight, it appears quite consequential. The difficulty in applying it is
discussed after the proof.
\begin{prop}
\label{prop:progress-bad}
Assuming that $(\tgt{v},\rho_0) \not\in \zero$,
\begin{equation}
\frac{d}{d\lambda}\xs(\tgt{v},{\rho}_\lambda)\Big|_{\lambda=0} < 0,
\label{eq:progress-1st-try}
\end{equation}
whenever the derivative exists. 
\end{prop}
\begin{proof}[Proof of Prop.~\ref{prop:progress-bad}]
Apply monotonicity (\ref{eq:monotonicity}) of $\xs_0$ to the two points
$(v_0,\rho_0), (v_1,\rho_1)\in \zero_0$ (as illustrated in Fig.~\ref{fig:bivariate})
to obtain
\begin{equation}
  \pair{\Rsd {v}}{\rho_1-\rho_0} < 0.
\label{eq:mono 2}
\end{equation}
The inequality is strict for the following reason.
If $(v_1,\rho_0)\in \zero_0$, then $(v_1 + \slct{\ldiff{\Phi}(\rho_0)},\rho_0)\in\zero$.
  However, since $v_1 = {v_0}+\Rsd v_0$, this says that $\rho_0$ is an interacting
  ground density for ${v_0}+\Rsd v_0 + \slct{\ldiff{\Phi}(\rho_0)} = \tgt{v}$,
  contrary to assumption.

Combining (\ref{eq:Rv-in-subdiff}) and (\ref{eq:mono 2}) shows that
the set \hbox{$\pair{\ldiff_\rho\xs(\tgt{v},\rho_0)}{\rho_1-\rho_0}$}
intersects $(-\infty,0)$.
On the other hand, if the derivative exists,
\begin{equation}
 \frac{d}{d\lambda}\xs(\tgt{v},\rho_0 +\lambda[\rho_1 - \rho_0])\Big|_{\lambda=0}
= \pair{w}{\rho_1 - \rho_0}
\end{equation}
for any $w \in {\ldiff_\rho\xs(\tgt{v},\rho_0)}$. The inequality (\ref{eq:mono 2}) follows.
\end{proof}

Unfortunately, there is a serious problem with this as a basis of a strategy.
To be able to use it in a non-blind way, we must be able to test the value of
$\xs(\tgt{v},{\rho}_\lambda) - \xs(\tgt{v},{\rho}_0)$.
As previously discussed, the only evident feasible way to do that is to obtain
${\rho}_\lambda$ as the second component of a point on $\zero_0$, which
means we need to know a potential having ${\rho}_\lambda$ as a ground density.

\subsection{A feasible strategy}\label{sec:feasible-strategy}

A second attempt to find a method of feasibly making progress
involves linear interpolation of the potential according to:
\begin{align}
  \label{eq:v-lambda}
v_\lambda
  &  = (1-\lambda) v_0 + \lambda v_1
    \nonumber \\
&  = v_0 + \lambda \Rsd v_0.
\end{align}
Corresponding densities $\rho_\lambda$ are defined implicitly via 
\begin{equation}
  \label{eq:nonlinear-rho-interpolation}
({v_\lambda},\rho_\lambda) = Z_0 v_\lambda.
\end{equation}
Caution: we are recycling notation here!
Although $\rho_\lambda$ interpolates between $\rho_0$ and $\rho_1$,
this interpolation is generally nonlinear, unlike in (\ref{eq:rho-interp}).

\begin{prop}
\label{prop:progress-good}
\begin{align}
  \inc(\tgt{v};\rho_\lambda,\rho_0)
  =
& \xs(\wh{v_0},\rho_\lambda)
    \nonumber \\
  & -\frac{1}{\lambda} 
 \Big[ \xs_0({v_\lambda},\rho_0) + \xs_0({v_0},\rho_\lambda) \Big].
\label{eq:prog-equals}
\end{align}
This is bounded above by either of the following:
\begin{subequations}
\begin{align}
& \lambda^{-1}\pair{{v_\lambda - v_0 }}{\rho_\lambda - \rho_0}  
-\pair{\wh{v_\lambda} - \wh{v_0}}{\rho_\lambda - \rho_0},
\label{eq:bound-a}
\\
&
\pair{(1-\lambda)\Rsd v_0 + \slct{\ldiff\Phi(\rho_0)} - \slct{\ldiff\Phi(\rho_\lambda)}}
{\rho_\lambda - \rho_0}.
\label{eq:bound-b}
\end{align}
\end{subequations}
\end{prop}
\begin{cor}
\label{cor:progress}
With the preceding notation, assuming $\rho_\lambda$ exists and $\Rsd v \neq 0$, 
\begin{equation}
  \label{eq:Delta-xs-upper-bd}
\inc(\tgt{v};\rho_\lambda,\rho_0) 
< \xs(\wh{v_0},\rho_\lambda) -\frac{1}{\lambda} \xs_0({v_0},\rho_\lambda).
\end{equation}
\end{cor}
Recall that $\xs_0$ and $\xs$ are everywhere non-negative.
The remarkable, and encouraging, aspect of the inequality
(\ref{eq:Delta-xs-upper-bd}) is the extra factor $\lambda^{-1}$ in
the negative term; more about this in the next subsection.
\begin{proof}[Proof of Prop.~\ref{prop:progress-good}]
Apply the identity (\ref{eq:v-shift}) to the expression
\begin{equation}
  \nonumber
\xs(\tgt{v},\rho_\lambda) - \xs(\tgt{v},\rho_0)  + \xs(\wh{v_\lambda},\rho_0)
\end{equation}
three times, replacing $(v,\rho,v')$ successively by 
$(\wh{v_\lambda},\rho_\lambda,\tgt{v})$, $(\wh{v_0},\rho_0,\tgt{v})$,
and $(\wh{v_0},\rho_0,\wh{v_\lambda})$.
This relies on $(\wh{v_\lambda},\rho_\lambda)\in\zero$.
In the resulting expression, each of $E(\wh{v_\lambda})$, $E(\tgt{v})$, and
$E(\wh{v_0})$ occurs once with a plus and once with a minus sign,
cancelling to leave
\begin{align}
\label{eq:step1}
\xs(\tgt{v},\rho_\lambda) - 
  \xs(\tgt{v},\rho_0)
= & \, - \xs(\wh{v_\lambda},\rho_0)
    \nonumber \\
& + \pair{\tgt{v}-\wh{v_\lambda}}{\rho_\lambda-\rho_0}.
\end{align}
Now, $\tgt{v} = \wh{v_0} + \Rsd v_0$ (definition of $\Rsd$).
  By (\ref{eq:v-lambda}), this is
  \hbox{$\wh{v_0} + \tfrac{1}{\lambda}(v_\lambda - v_0)$}.
Substitute for ${\tgt{v}}$ in the RHS of (\ref{eq:step1}), reducing it to
\begin{equation}
\nonumber  
- \xs(\wh{v_\lambda},\rho_0) +
\frac{1}{\lambda} \pair{v_\lambda - v_0}{\rho_\lambda - \rho_0}
     - \pair{\wh{v_\lambda} - \wh{v_0}}{\rho_\lambda-\rho_0}.
\end{equation}
Drop the negative first term here to obtain the upper bound (\ref{eq:bound-a}).
The second form (\ref{eq:bound-b}) of the upper bound follows upon the substitution
\hbox{$\wh{v_\lambda} - \wh{v_0} = {v_\lambda} - {v_0} +
 \slct{\ldiff\Phi(\rho_0)} - \slct{\ldiff\Phi(\rho_\lambda)}$}.

Returning to the previous display, use the cross-difference identity
(\ref{eq:cross-difference}) once for the reference system and once for the
perturbed system to rewrite that display as
\begin{equation}
  \nonumber
  -  \frac{1}{\lambda}\Big[
    \xs_0(v_\lambda,\rho_0)  + \xs_0(v_0,\rho_\lambda) \Big]
    + \xs(\wh{v_0},\rho_\lambda).
\end{equation}
Equating to the LHS of (\ref{eq:step1}) yields (\ref{eq:prog-equals}).
\end{proof}

\subsection{Analyticity}\label{sec:analyticity}

The question now is, under what circumstances is the RHS of the inequality
(\ref{eq:Delta-xs-upper-bd}) negative for some range of $\lambda$?
If both $\xs(\wh{v_0},\rho_\lambda)$ and $\xs_0({v_0},\rho_\lambda)$
are $\mathcal{O}(\lambda^2)$, that would be more than enough.
Recall that $v_\lambda = (1-\lambda) v_0 + \lambda v_1$ and
$\rho_\lambda$ is \emph{a} noninteracting ground density for
$v_\lambda$.
Both $\xs(\wh{v_0},\rho_\lambda)$ and $\xs_0({v_0},\rho_\lambda)$ certainly
have a minimum (zero) at $\lambda=0$.
If $\rho_\lambda$ varies at all smoothly, we would expect both these
excess energies to be quadratic in $\lambda$ near the minimum, exactly
as needed.

Supposing $F$ is an exact functional, so that $\xs$ comes from a well-defined
quantum mechanical problem, the following can be proved\cite{Lammert-21}:
If the noninteracting problem for $v_0$, and the interacting problem for
$\wh{v_0}$, both have a nondegenerate ground state with nonzero spectral gap,
then both these excess energies are not just $\mathcal{O}(\lambda^2)$,
but \emph{analytic} at $\lambda=0$.
On the other hand, if the nondegeneracy and gap conditions are not 
satisfied, we should not be at all surprised if the excess energies
behave in a way which dashes our hopes. 
The strategy of section \ref{sec:feasible-strategy} is
therefore conditionally vindicated.

\section{Interlude: toward topology}\label{sec:interlude-top}

The rest of this paper develops a functional analytic picture
which is not directly dependent on our analysis of Kohn-Sham machines,
but very much influenced by it.
Questions asked, and hypotheses imposed are chosen to be relevant.
For instance, in asking about limits of a sequence $((v_n,\rho_n))$
of ground pairs, we will decide it is reasonable to ask that the
$F(\rho_n)$ be bounded on the grounds that the KS machine provides this
information when it generates a ground pair.

We saw that the course of a Kohn-Sham computation can be distilled
into a sequence $(v_n,\rho_n)$ of ground pairs.
Moreover, barring exceptions, the computation can be done such that
$\xs(\tgt{v},\rho_{n+1}) < \xs(\tgt{v},\rho_{n})$. This we called
``progress'', but maybe we should call it $\xs$-progress as there may
be other sorts.
Since we do not know any ground densities of $\tgt{v}$ (else we would not
be doing the computation), deciding whether $\rho_{n+1}$ is closer to such
than is $\rho_n$ certainly cannot be done directly, at least.
Surely, though, we could see whether $v_{n+1}$ is closer to $\tgt{v}$ than
$v_n$? Only if we know what ``closer'' means.
If we had a metric $d'$ on $\SV$, that would provide one answer, and
we could speak of ``$d'$-progress''.

This brings us to the issue of topologies on our function spaces, which
we have so far deliberately avoided.
The next section contains a review of relevant ideas, tailored to our needs.
For us, equipping $\SV$ and $\SD$ with topologies is not merely a matter of
mathematical convenience, but has physical significance, and will
be done based on the considerations of section \ref{sec:to-DFT}.
After all, how do we distinguish one state (i.e., density) from another?
By finding an observable which takes differing values for them.
Thus arises the most physically-grounded notion of \textit{neighborhood}
of a density.

\section{Topological notions}\label{sec:topology}

This section review some important topological concepts and
relates them to the physical state-observable duality.
Because of the latter, readers already comfortable with all the
mathematics maybe should skim it.
By \textit{topology}, I refer to the classical idea of
defining neighborhoods of points in a point set, closely
related to approximation.
Actually, we do not deal with general topologies, but metrics and semimetrics.
One may wonder whether even that is excessive.
For that reason, it bears emphasizing at the outset that
we will do this in order to ground the mathematics physically.
Following the development in section \ref{sec:QM-to-DFT},
the fundamental means at our disposal to distinguish densities and
define neighborhoods is via the observables. There are infinitely many
of these, and they naturally give a system of \textit{seminorms}.
If we choose to work with a norm, for convenience, it is desirable that
it have some justification tracing back to the observables.

\subsection{Metrics, norms, semimetrics, seminorms}\label{sec:metrics}

A metric on a set $X$ is a map $\Arr{X\times X}{d}{[0,\infty)}$ (distance function)
satisfying
\begin{enumerate}
\item[] $d(x,y) = d(y,x)$ \hfill (symmetry)
\item[] $d(x,z) \le d(x,y) + d(y,z)$   \hfill (triangle inequality)
\item[] $d(x,y) > 0 \;\Rightarrow\; x \neq y$ 
\item[] $d(x,y) = 0 \;\Rightarrow\; x = y$ 
\end{enumerate}
The set together with the metric, $(X,d)$ is a \textit{metric space}.
The \textit{open ball} of radius $r$ about $x\in X$ is the set
\begin{equation}
  \nonumber
B(r;x) \defeq \setof{y\in X}{d(x,y) < r}
\end{equation}
of points at distance less than $r$ from $x$.
If $d$ and $d'$ are two metrics on the same space, $d'$ is \textit{stronger}
than $d$, written $d \precsim d'$, or $d' \succsim d$, if for every $r>0$, there
is $r'>0$ such that $B'(r',x) \subseteq B(r,x)$ for \emph{every} $x$.
Here, $B'$ denotes an open ball for $d'$.
$d$ and $d'$ are \textit{equivalent}, $d \sim d'$,
in case both $d\precsim d'$ and $d' \precsim d$.
These comparisons are significant for convergence of sequences.
Three ways to express the same thing are:
sequence $(x_n)$ converges to $x$ with respect to $d$,
$\lim_{n\to\infty} d(x,x_n) = 0$,
and, for any $r>0$, some tail of the sequence is inside $B(r,x)$.
Hence, $d \precsim d'$ implies that every $d'$-convergent sequence
is \hbox{$d$-convergent}.

If $X$ is a vector space, metrics which are compatible with the linear
structure are of most interest. This means
$d(x+z,y+z) = d(x,y)$ (translation invariance) and, for $c\in\Real$,
$d(cx,cy) = |c| d(x,y)$ (homogeneity).
A corresponding \textit{norm} can then be defined as the distance
$\|x\| = d(0,x)$ from the origin.
Such a metric is recovered from the corresponding norm as
$d(x,y) = \|x-y\|$.

The last two listed defining conditions for a metric pertain to its
role in distinguishing points.
The third condition shows how it does that, and the fourth,
\textit{separation},
says that the metric can distinguish any distinct points.
Dropping the separation condition yields the definition of a
\textit{semimetric}. A single semimetric may fail to separate
points, but a collection $\setof{d_i}{i\in \mathcal{I}}$ of
semimetrics can \textit{collectively} separate, even if none
does so individually. That is, for each $x\neq y$, there is
some $i\in \mathcal{I}$ such that $d_i(x,y) > 0$.
A sequence $(x_n)$ converges to $x$ with respect to the system of seminorms
$\setof{d_i}{i\in \mathcal{I}}$ if and only if $d_i(x_n,x)\to 0$ for each $i$.
Extending our comparison ($\precsim$) to systems of semimetrics has a slight
subtlety. One way to proceed is to use open balls again.
The ``size'' of the open ball $B(J,r;x) = \setof{y}{d_i(y,x) < r, \forall i\in J}$
is parameterized by not only a radius, but also a selection ($J\subset \mathcal{I}$) of a
\textit{finite} number of seminorms.
Then, $\setof{d_i}{i\in\mathcal{I}} \precsim \setof{d'_j}{j\in\mathcal{I}'}$
if and only if for any $d$ size $(I,r)$, there is a $d'$ size $(I',r')$ such that
$B'(I',r';x) \subseteq B(I,r;x)$ for every $x$.
As concerns convergence, a collectively separating \emph{finite} system
$\{d_1,\ldots,d_n\}$ can be replaced by the single metric
$d_*(x,y) = \max(d_1(x,y),\ldots,d_n(x,y))$. Hence, only infinite systems of
semimetrics are really of interest.

Just as for the passage from metric to norm,
to make a seminorm respect the linear structure of a vector space, 
one imposes translation invariance and homogeneity.
Such a compatible semimetric is a \textit{seminorm}.
(Terminological note: \textit{seminorm} is standard.
Accepting that, \textit{semimetric} seems natural. However, 
what we are calling \textit{semimetric} is called \textit{pseudometric}
by some.)

\subsection{Seminorms and dual pairs}\label{sec:seminorm-dual}

Seminorms have been lurking all along in our pairing maps.
Suppose $\SX$ and $\SV$ form a dual system (Def.~\ref{def:dual-system}).
Each $v\in\SV$ defines a proper seminorm $p_v$ on $\SX$, defined by
\begin{equation}
  \label{eq:pv-seminorm}
x \mapsto p_v(x) \defeq |\langle{v,x}\rangle|.  
\end{equation}
Similarly, each $x\in\SX$ defines a seminorm on $\SV$.
No single $p_v$ separates, but the entire system of seminorms separates
collectively.
For $\SX = \vecsp\SD$ and $\SV$ our spaces of states and observables, respectively,
this is something we should insist on.
If two states cannot be distinguished by \textit{any} observable, on what ground would
we say they are distinct? Relatedly, if we admit the physical meaningfulness
of a set of observables, it is very unclear on what grounds we could
reject the physical meaningfulness of the corresponding seminorms and
the topology which they generate. 
Since systems of seminorms arising this way are of great importance to us,
we introduce a notation. 
\begin{defn}
For a dual pair $\pair{\SV}{\SX}$,
the system $\setof{p_v}{v\in\SV}$ of seminorms defined in (\ref{eq:pv-seminorm})
is denoted $\sigma(\SX,\SV)$. Swapping the roles of $\SX$ and $\SV$ gives the
system $\sigma(\SV,\SX)$ $\SV$.
If $\SD$ is a subspace of $\SX$, we write $\sigma(\SD,\SV)$ for the system of semimetrics
induced from $\sigma(\SX,\SV)$.
\end{defn}

\subsection{Norm compatibility with a dual system}\label{sec:norm-compatibility}

Once $\SX$ has a topology, we have a new criterion with which to
distinguish linear functionals, namely, those which are continuous.
It turns out that the linear functionals on $\SX$ continuous with respect
to $\sigma(\SX,\SV)$ are those (and only those) of the form
$x \mapsto \pair{v}{x}$ for $v\in\SV$.
Physically, this makes sense: the linear observables ought to be exactly the
continuous linear functionals on states, or something has been chosen incorrectly.

It is not as easy to work with a seminorm system such as $\sigma(\SX,\SV)$
as with a simple norm, at either the level of general results or that of specific
spaces. This motivates us to equip $\SX$ with a norm, but it also raises
the question of potential grounds for considering a norm to be ``physical''.
I propose a principle based on the observation of the previous paragraph.
A topology $\tau$ on $\SX$, defined by seminorms, is said to
be \textit{compatible} with the duality $\pair{\SV}{\SX}$ if
the set of linear functionals on $\SX$ which are continuous with respect
to $\tau$ are exactly those of the form $x \mapsto \pair{v}{x}$ for $v\in\SV$.
Then, the principle is that, to the extent that the choice $\SV$ of
observables is physical, topologies compatible with the duality $\pair{\SV}{\SX}$
are the ``more physical'' ones.

This matter of topologies compatible with a given duality is a standard
chapter of the theory of locally convex spaces.
(Often literally, e.g., Chapter III of Horv\'{a}th's book\cite{Horvath}.)
We list some relevant facts.
Not only do all topologies compatible with a given duality have the same
  continuous linear functionals, but also
  (i) the same lower semicontinuous convex functions into $\overline{\Real}$,
  (ii) the same closed convex subsets of $\SX$, (iii) the same bounded sets.

An important observation is that, if there is a norm on $\SX$ compatible with
the duality, it is essentially unique, and defined by the weakest seminorm dominating all
the $p_v$ for $v\in\SV$.
Fortunately, we have such a case.
\hbox{$\SV = \mathcal{L}^\infty(\mathcal{M}) + \mathcal{L}^{3/2}(\mathcal{M})$}
continues to be the dual space of $\vecsp \SD$ when the 
system $\sigma(\SV,\vecsp \SD)$ is stengthened to the norm 
\begin{equation}
\label{eq:L1+L3}
\|x\| = \|x\|_{1 \cap 3} \defeq \|x\|_1 + \|x\|_3.    
\end{equation}
$\vecsp\SD$ is \textit{not} a Banach space under this norm.
Its \textit{completion} (see section \ref{sec:completeness})
is the Banach space \hbox{${L}^1(\mathcal{M})\cap{L}^3(\mathcal{M})$}.

With the canonical norm
\begin{equation}
  \label{eq:SV-norm}
\|v\|' \defeq \sup \setof{|\pair{v}{x}|}{x\in\vecsp\SD, \, \|x\| = 1},
\end{equation}
$\SV$ becomes a Banach space.
A norm which is equivalent to this canonical one, and
possibly more convenient, or at least more explicit, is
\begin{equation}
  \label{eq:infty+3/2-norm}
\|v\|_{\infty+\frac{3}{2}} \defeq \inf\setof{\|v_1\|_\infty + \|v_2\|_{\frac{3}{2}}}{v = v_1 + v_2}.  
\end{equation}
However, we will not actually make any use of this concrete form.

\subsection{Variations on continuity}\label{sec:continuity}

We collect some variations on the concept of continuity for metric spaces.
Recall that a function $\Arr{\SX}{f}{\SY}$ between metric spaces
is continuous at $x\in\SX$ iff, given $\delta > 0$,
there is an $\epsilon > 0$ such that $f$ carries the ball of radius
$\epsilon$ centered at $x$ \textit{into} the ball of radius $\delta$
centered at $f(x)$.

In section \ref{sec:RCI}, we shall use a slightly stronger form of continuity,
as follows. 
\begin{defn}[locally Lipschitz continuous]\label{def:L-cts}
For a metric space $\SX$, a function $\Arr{\SX}{f}{\Real}$
is \textit{locally Lipschitz continuous}
(for short, \textit{locally L-continuous})
\defined for each point $x\in \SX$, there is a neighborhood $U \ni x$ and $K > 0$ such that
\begin{equation}
y,z \in U \;\Rightarrow\; |f(y)-f(z)| < K d(y,z).
\end{equation}
\end{defn}
\smallskip
Example: the function $x\mapsto \sqrt{|x|}$ is
locally \hbox{L-continuous} on $\Real\setminus \{0\}$, but only
continuous at zero.

Just as for the unilateral forms of derivative
introduced in Def.~\ref{def:semiderivative}, a unilateral form of continuity is
relevant in optimization situations.
\begin{defn}[lower/upper semicontinuity]\label{def:lsc}
  A function $\Arr{\SX}{f}{\overline\Real}$ on a topological space $\SX$ is
  \textit{lower semicontinuous} (lsc) at $x\in\SX$ when, for any
  level $c < f(x)$, there is a neighborhood $U$ of $x$ such that
  \begin{equation}
    \nonumber
 y\in U \;\Rightarrow\; f(y) > c.
  \end{equation}
  \textit{Lower semicontinuous} without qualifier means lsc everywhere.
  $f$ is \textit{upper semicontinuous} (usc) if $-f$ is lsc.
\end{defn}

For a convergent sequence $x_n \to x$, lower semicontinuity of $f$ implies
that $\liminf_{n\to\infty} f(x_n) \ge f(x)$. The value of $f$ at the limit point
$x$ might be ``smaller than anticipated'', but not ``larger than anticipated''.
A real-valued function is continuous at a point iff it is both lsc and usc there.
The concept of lower semicontinuity is very important for us because $F$ is lsc,
but not usc (see section \ref{sec:F-not-usc}).

If $S$ is a set of lsc functions, then their pointwise supremum,
$f(x) \defeq \sup \setof{g(x)}{g\in S}$ is also lsc.
In particular, if $S$ consists of continuous functions, then the
supremum is lsc, though there is no reason, in general to suppose it continuous.
For an pertinent example, consider $E$. It is the pointwise infimum of affine
functionals $v \mapsto F(\rho) + \pair{v}{\rho}$, hence is usc if those are
continuous, which they will be if $\SV$ is equipped with a system of
seminorms at least as strong as $\sigma(\SV,\SD)$. 

Finally, we introduce a weakening of continuity which will be useful because
it allows us to bound how discontinuous $F$ can be in certain circumstances.
\begin{defn}[almost continuous]\label{def:almost-cts}
  $\Arr{\SX}{f}{\Real}$ is
  
\begin{enumerate}
\item 
{\textit{$\epsilon$-almost continuous at $x$}}
precisely if:
\hfill\newline
for any $\epsilon' > 0$, there is $\delta$ such that 
\begin{equation}
d(y,x) < \delta \;\Rightarrow\;  |f(x) - f(y)| < \epsilon + \epsilon'
\end{equation}
\item {$\epsilon$-almost continuous} \textit{on} $\SX$ 
precisely if:
\hfill\newline
  $f$ is {$\epsilon$-almost continuous} at $x$ for every $x\in\SX$.
\item
{$g$-almost continuous}, where $\Arr{\SX}{g}{[0,\infty)}$, precisely if:
for every $\epsilon > 0$, $f$ is $\epsilon$-almost continuous on $\{g \le \epsilon\}$
\end{enumerate}
\end{defn}

\subsection{Complete metric spaces}\label{sec:completeness}

Due to its importance in the investigation, we conclude this section with a brief review
of the concept of completeness for metric spaces.
Roughly, a metric space is \textit{complete} if a sequence
actually has a limit whenever it ``appears to be converging'' in the following sense.
\begin{defn}[Cauchy]
The sequence $(x_n)\subseteq \SX$ is \textit{Cauchy} \defined 
  \begin{equation}
\mathrm{diam}\setof{x_n}{n\ge N} \to 0 \text{ as } N\to\infty.    
\end{equation}
\end{defn}
The \textit{diameter} of a set $A$, $\mathrm{diam}\, A$
is $\sup\setof{d(x,y)}{x,y\in A}$.
\begin{defn}[Complete]
  \label{defn:complete}
  A metric space $\mathscr{X}$ is \textit{complete} if every Cauchy sequence
  in $\mathscr{X}$ has a limit in $\mathscr{X}$.
\end{defn}
For a familiar example of a metric space which is \textit{not} complete,
consider the rational numbers $\Rat$ with the ordinary distance function.
If $x_n$ is $\sqrt{2}$ to $n$ decimal places, then $(x_n)$ is Cauchy, but
does not converge to anything since $\sqrt{2}$ is not in $\Rat$.
A Banach space is a complete normed space. There is a canonical, abstract,
way to complete any normed space $V$. The completion is a Banach space and
$V$ is dense in it.
Then, any Cauchy sequence has a limit in the completion.
This seems very convenient, but is not always appropriate.
Later we will be interested in the metric $d_1$ on the space of densities $\SD$
which derives from the $L^1$ norm. We will not use a completion because we will
want to know that limits are in $\SD$ itself.

Our partial order on metrics, $\precsim$, behaves well with respect to
completeness. Namely, if $(X,d)$ is complete and \hbox{$d\precsim d'$}, then
$(X,d')$ is also complete. 

\section{Interlude: general strategy}\label{sec:interlude-2}

Suppose we have found a well-motivated metric on $\SV\times\SD$,
and return to the sequence $((v_n,\rho_n))$ of ground pairs.
Questions which naturally arise are:
Does it converge if it is Cauchy (see section \ref{sec:completeness} for this notion)?
If it does, is the limit a ground pair?
The following sections put together a topological perspective on DFT.
We consider regularity properties of $E$ and $F$, the relation between
the energetic version of nearly a ground pair (small excess energy) and
distance to $\zero_0$ or $\zero$, and convergence of sequences of
ground pairs. 

\subsection{Room for error}

Our analysis of Kohn-Sham machines assumed that they produce points
exactly on $\zero$, i.e., with zero excess energy $\xs$.
Assuming that only $\xs(v_n,\rho_n) < \epsilon$ is an idealized model
of a certain kind of error.
In the following sections, therefore, we will be interested not only  in
$\zero$, but also sets of bounded $\xs$ in $\SV\times\SD$, in order to
understand how the conclusions are robust against such error.

\subsection{the axiomatic approach}

Additional axioms will be added to \ref{A:D} -- \ref{A:dual-pair}
already announced. They will be \emph{motivated} by what we can deduce
about the exact quantum mechanical situation, but are expressed at
the DFT level. This style of working allows us to keep track of exactly
what we have used from the underlying QM (not a lot), and gives room for
the results to apply to model functionals.

There will only ever be a single $F$ involved.
However, it need not be an exact functional, clearly traceable to
a quantum mechanical Hamiltonian.
Any $F$ which satisfies the axioms will do, so
it could be $F_0$, an exact $F$, or $F_0 + \Phi$ for a model HXC energy.
The axioms reflect properties of exact functionals, but
are not particularly constraining.

Final results are funnelled through the axioms, so to speak.
There is work to be done both in proving that the axioms are satisfied in
standard interpretation, and in getting from them to claims formally
stated as theorems.
This is not always most efficient approach.
Two later axioms will supercede earlier ones. Choice of axioms aims
for mathematical simplicity, physical transparency, and
generality (hence flexibility in application). 

\subsection{$F$ is very far from continuous}\label{sec:F-not-usc}

In the physics literature, it is often implicitly assumed that intrinsic
energy $F$ is well-behaved, continuous at least, and possibly smooth.
This is not only unjustified, but incorrect.
With respect to the norm $\|\cdot\|$ already mentioned, and dealt with
in the next section, the exact functional $F$ is lsc, but \emph{not}
usc. In fact, $F_0$ already has this problem.
To see this, consider a density $\rho$, and select a region $U$ and $\epsilon > 0$.
By adding oscillations of bounded amplitude but increasingly small
wavelength to $\rho$ in the region $U$, we can produce a sequence of
densities $\rho_n$ such that $\|\rho_n-\rho\| < \epsilon$, but $F_0(\rho_n) > n$.
(See Section II of \cite{Lammert10a} for further discussion.)
Hence, $F_0$ is \emph{unbounded above on every neighborhood}.
The excess energy $\xs$ inevitably inherits this problem.
This is worth emphasizing because some of what follows,
though by no means all, 
would be somewhat trivial if $F$ were continuous.
In addition, we also have $E$ and $\xs$ to worry about.

%
%
%
%

\section{Structure and regularity I}\label{sec:RCI}

\subsection{New postulates}\label{sec:RCI-postulates}

In addition to \ref{A:D} -- \ref{A:dual-pair} from section~\ref{sec:KS-assumptions}, we
now assume

\begin{enumerate}[label=B{\arabic*}.,series=baxs,ref=B\arabic*]
\item \label{B:dom-E}
$\dom E \supseteq \SV$.
\item \label{B:norm}
$\SV$ is the topological dual of $(\vecsp\SD,\|\cdot\|)$
with respect to the pairing $\pair{}{}$.
\item \label{B:lsc}
$F$, extended to the completion of \hbox{$(\vecsp\SD,\|\cdot\|)$} by
$F\equiv +\infty$ off $\SD$, is lower semicontinuous.
\end{enumerate}

Recall that `$\dom$' indicates the set on which a function takes
a proper (noninfinite) value.
All the potentials under consideration are in $\SV$, so \ref{B:dom-E}
might reasonably have been written with ``$=$'' in place of ``$\supseteq$''.
This version makes the point that it would not be a problem if $E$ were
well-defined and finite for something outside $\SV$.
With \ref{B:dom-E}, all our functions $F$, $E$, and $\xs$ take proper values over
all of $\SV\times\SD$.
Axiom \ref{B:lsc} is perhaps somewhat unsatisfactory insofar as it is not immediately
clear what property of $F$ as given implies that such an extension is possible, and
one would prefer not to have to think outside $\SD$, or $\vecsp \SD$. This will be
addressed in Section \ref{sec:cmp-lsc}. For now we work with this fairly standard
form.

Together with the pairing $\pair{\cdot}{\cdot}$, the
norm $\|\cdot\|$ on $\vecsp \SD$ induces a canonical norm
(\ref{eq:SV-norm}) $\|\cdot\|'$ on $\SV$, under which it is a Banach space.
The corresponding metrics are denoted by $d$ and $d'$, respectively.

\subsection{Standard interpretation}

The interpretation is
[See Eqs. (\ref{eq:L1+L3}) and (\ref{eq:infty+3/2-norm})]
\begin{align}
& \| x \| \defeq \|x\|_{1\cap 3}
\nonumber \\
  & \|v\|' \defeq \|v\|_{\infty+\frac{3}{2}},
    \nonumber \\
& \pair{v}{x} = \int_{\Real^3} v(x)\rho(x)\, dx
\nonumber
\end{align}
The pairing was already defined on a bigger set than $\SV\times\SD$,
so the extension described is not really necessary, but it is worth noting
that the extension recovers the original pairing on the bigger set.

\subsection{Structure theorem}\label{sec:RCI-result}

In this section, we equip $\SV\times\SD$ with the metric
\begin{equation}\label{eq:product-metric}
(d'+d)((v,\rho),(v',\rho')) \defeq d'(v,v') + d(\rho,\rho').  
\end{equation}
Until further notice, convergence will be considered with respect to
$d'$, $d$ and $d'+d$ in $\SV$, $\SD$, and $\SV\times\SD$, respectively.

A subset of $\SV\times\SD$ over which $\xs$ is
bounded (i.e., a subset of $\{\xs \le M\}$ for some $M < \infty$)
is called a \hbox{\emph{$\xs$-bounded set}}.
Later we will be interested in \hbox{\emph{$F$-bounded}} sets, which are defined
similarly. In that case, however, there is an ambiguity:
\hbox{$\{F\le M\}$} could be a subset of $\SD$, or a subset of $\SV\times\SD$
with unrestricted $\SV$ coordinate.
Context will make clear which is intended.

$F$ is lsc by assumption (\ref{B:lsc}), while $E$ is usc by
construction (see section \ref{sec:continuity}).
$\xs$ is then the sum of lower semicontinuous functions of density, $F(\rho)$, 
of potential, $-E(v)$, and a separately continuous function $(v,\rho)\mapsto \pair{v}{\rho}$.
$\xs$ is therefore separately lsc. Just as for continuity, \emph{joint} lower semicontinuity
(i.e., as a function on $\SV\times\SD$) is not in general a consequence of separate lower
semicontinuity.
Much of the force of the following Proposition~\ref{prop:RCI} is in showing
that the situation is actually better than just observed.
The improvement is clear as regards $E$ (conclusion 2).
Conclusion 1, although stated in a somewhat raw form,
implies that $\xs$ is lsc on $\SV\times\SD$,
as is thoroughly explained in Section \ref{sec:cmp-lsc}.
Conclusions 3 and 4 show that $F$ is better behaved in restriction to subsets of
small excess energy.
Beware of misinterpretation.
Conclusion 3 does \emph{not} mean that $F$ is continuous with respect to $d$ on
the set of v-representable densities.
Rather, we can rephrase it as:
$F(\rho')$ is close to $F(\rho)$ if $\rho'$ is close to $\rho$
\textit{and} a realizing potential for $\rho'$ is close to one for $\rho$.
The relevance of considering $\zero$ is that Kohn-Sham computation delivers
points on $\zero$, or, in a less-idealized version, on $\{\xs\le\epsilon\}$.

\begin{prop}\label{prop:RCI}
Assume \hbox{\ref{A:D} -- \ref{A:dual-pair}, \ref{B:dom-E}} -- \ref{B:lsc}.
Then, on \hbox{$(\SV\times\SD,d'+d)$},
\begin{enumerate}
\item 
For $\epsilon < \infty$, $\{\xs \le \epsilon\}$ is complete.
\item 
$E$ is locally L-continuous
\item
$F$ is locally L-continuous on $\zero$
\item
$F$ is $\xs$-almost continuous
\end{enumerate}
\end{prop}
In more concrete terms directly related to Kohn-Sham computation,
Prop.~\ref{prop:RCI} has the following immediate consequence.
Suppose
\begin{itemize}
\item 
$((v_n,\rho_n))\in\zero$
\item
$(v_n,\rho_n) \to (v,\rho)$
\end{itemize}
Then,
\begin{itemize}
\item
$(v,\rho) \in \zero $
\item
$F(\rho) = \lim F(\rho_n)$
\item $E(v) = \lim E(v_n)$
\end{itemize}


\subsection{\ref{B:dom-E} -- \ref{B:lsc} hold in standard interpretation}\label{sec:RCI-demos}
 
\begin{proof}[Proof of \ref{B:dom-E}]
  The problem is to show that, for fixed $v$,
  $F(\rho) + \pair{v}{\rho}$ is bounded below with respect to $\rho$.
  The crucial facts are (i) $\|\rho\|_3 \le a + b F(\rho)$, where $b > 0$,
  and (ii) $v$ can be split as $v' + v''$, where $v'\in \mathcal{L}^1$ and
  $\|v''\|_{3/2}$ is as small as desired.
The second item is Lemma \ref{lem:v-is-nearly-L-infty} below.
Using normalization of $\rho$ ($\mathcal{N}$ particles),
\begin{equation}
  \nonumber
  F(\rho) + \pair{v}{\rho} \ge F(\rho)
  - \mathcal{N}\|v'\|_\infty - \|v''\|_{3/2}(a + b F(\rho)).
\end{equation}
Now, it is only necessary to choose $v''$ so that \hbox{$1 - \|v''\|b > 0$}.
\end{proof}
Actually, the first ``crucial fact'' here is the reason\cite{Lieb83}
for choosing the $L^3$ norm.
\begin{lem}\label{lem:v-is-nearly-L-infty}
For any $\epsilon > 0$, 
\hbox{$v\in\mathcal{L}^\infty(\Real^3) + \mathcal{L}^{3/2}(\Real^3)$}
admits a splitting $v = v' + v''$, with
  $v' \in\mathcal{L}^\infty(\Real^3)$ and
  $\| v''\|_{3/2} < \epsilon$.
\end{lem}
\begin{proof}
For $m > 0$, split $v$ as $v_m + v^m$, with
  \begin{equation}
    v_m \defeq
    \begin{cases}
      -m & v < -m \\
      v & -m \le v \le m \\
      m & v > m
    \end{cases}.
  \end{equation}
$v_m \in \mathcal{L}^\infty(\Real^3)$ and
$v^m \in \mathcal{L}^{3/2}(\Real^3)$,
while $v^m \to 0$ pointwise almost everywhere
as $m\to\infty$. Hence, by the Dominated Convergence Theorem,
$\|v^m\|_{3/2} \to 0$ as $m\to\infty$.
\end{proof}
\begin{proof}[Proof of \ref{B:norm}]
  $\vecsp \SD$ is dense in the Banach space \hbox{$L^1\cap L^3$}, the
  topological dual of which is the Banach space \hbox{$L^{\infty}+L^{3/2}$}.
Refer to discussion in section~\ref{sec:norm-compatibility}.
\end{proof}
 \begin{proof}[Proof of \ref{B:lsc}]
See Thm. 3.6 of Ref. \onlinecite{Lieb83}.
\end{proof}

\subsection{Proof of Proposition \ref{prop:RCI}}\label{sec:RCI-proof}

\noindent A. {$E$ is locally L-continuous}.

\begin{proof}
This follows from \ref{B:dom-E}, using 1.4.1 and 1.7.4 of Schirotzek\cite{Schirotzek}.  
\end{proof}

\medskip\noindent B. $(v,\rho)\mapsto \pair{v}{\rho}$ is locally L-continuous.

\begin{proof}
  From
\begin{align}
\pair{v}{\rho}-\pair{\tilde{v}}{\tilde\rho} =
   \pair{v-\tilde{v}}{\rho-\tilde\rho}
                                  \nonumber \\
   +    \pair{v-\tilde{v}}{\tilde\rho}
    +    \pair{\tilde{v}}{\rho-\tilde\rho}
                                  \nonumber
\end{align}
deduce
\begin{align}
|\pair{v}{\rho}-\pair{\tilde{v}}{\tilde\rho}| \le
&    \|v-\tilde{v}\|' \|\rho-\tilde\rho\|
                                  \nonumber \\
  +    \|{v-\tilde{v}}\|' \|{\tilde\rho}\|
&    +    \|{\tilde{v}}\|' \|{\rho-\tilde\rho}\|.
                                  \nonumber
\end{align}
Considering, for example, the open set $\|v\|' < c$, $\|\rho\| < c$,
the RHS above can be bounded by $3c[ d'(v,\tilde{v}) + d(\rho,\tilde\rho)]$.
\end{proof}

\medskip\noindent C. $\{ \xs\le \epsilon \}$ is complete.

\begin{proof}
The Cauchy sequence $(v_n,\rho_n)\subset \{\xs\le\epsilon\}$
has a limit $(v,\rho)\in\SV\times\SX$ since $\SV$ and $\SX$ are
complete under $d'$ and $d$, respectively.
Now,  \hbox{$\xs(v,\rho) = F(\rho) + \pair{v}{\rho} - E(v)$}, and we need
to show that \hbox{$\xs(v,\rho) \le \liminf \xs(v_n,\rho_n)$}.
This follows because $F$ is lsc
(\ref{B:lsc}), while $E$ and $\pair{}{}$ are continuous, as shown in
items A and B.
\end{proof}

\medskip\noindent D. $F$ is locally L-continuous on $\zero$.

\begin{proof}
  Given the preceding, the proof of is simple.
  $F(\rho) = E(v) + \pair{v}{\rho} + \xs(v,\rho)$.
  The last term on the RHS is identically zero on $\zero$,
  while the first and second are locally
  L-continuous by items A and B, respectively.
\end{proof}

\medskip\noindent E.
$F$ is $\epsilon$-almost continous
on $\{\xs\le\epsilon\}$.

\begin{proof}
The first two terms on the RHS of $F(\rho) = E(v) + \pair{v}{\rho} + \xs(v,\rho)$
are continuous functions, by preceding results, while the final term is in the
interval $[0,\epsilon]$.
\end{proof}

\section{
Nearly-a-ground-pair versus
near-a-ground-pair}\label{sec:near-vs-nearly}

If $\xs(v,\rho)$ is small, then $(v,\rho)$ is \emph{nearly a ground pair}
in an obvious sense. But, does that imply that there is a
genuine ground pair nearby in $\SV\times\SD$?
The latter occurence, $(d+d')((v,\rho),\zero)$ is small, is naturally
described as $(v,\rho)$ is \emph{near a ground pair}.
This section is concerned with the extent to which these two concepts 
are commensurate.

All the axioms announced so far,
\hbox{ \ref{A:D} -- \ref{A:dual-pair} and \ref{B:dom-E} -- \ref{B:lsc}},
are assumed here.


\subsection{Nearly-a-ground-pair implies near-a-ground-pair}

The low (excess) energy part of the product space
\hbox{$\SV\times\SD$} is metrically close to the ground pairs $\zero$.

\begin{prop}
\label{prop:nearly-implies-near}
If $(v,\rho)$ is nearly a ground pair, i.e.,
$\xs({v},\rho)$ is small, then it is near some ground pair:
\begin{equation}
(d' + d)\Big(({v},\rho)\,,\,\zero\Big) < 2\sqrt{\xs(v,\rho)}
\end{equation}
\end{prop}
\begin{proof}
See Thm. I.6.2 (p. 31) of Ref. \onlinecite{ET},
Cor. 2 of \S 5.4 (p. 262) of Ref. \onlinecite{Aubin+Ekeland},
or Cor. 1.93 (p. 65) of Ref. \onlinecite{Penot}.
This is an application of the Ekeland variational principle\cite{Ekeland-74,Ekeland79}.
\end{proof}

\subsection{Perturbing the potential of a ground pair does not increase
excess energy much}
  
Nothing quite so straightforward or satisfactory is possible in the opposite direction. 
Here is why. We have
\begin{equation}
\nonumber
\xs({v},\rho) - \xs(v',\rho') =
[\xs({v},\rho) - \xs(v,\rho')] + [\xs({v},\rho') - \xs(v',\rho')].
\end{equation}
Now, Lemma \ref{lem:excess-is-Lipschitz} below shows that the
second bracketed term can be bounded as
\begin{equation}
\nonumber
|\xs({v},\rho') - \xs(v',\rho')| = (a + b F(\rho'))\|v-v'\|'.
\end{equation}
For the other bracketed term,
\begin{equation}
\nonumber
\xs({v},\rho) - \xs(v,\rho') = F(\rho) - F(\rho') + \pair{v}{\rho-\rho'}.
\end{equation}
The difference between $F(\rho)$ and $F(\rho')$ is uncontrollable, even as $\rho'\to\rho$,
because $F$ is unbounded above on every neighborhood. The best we can hope for is
that if $(v,\rho)$ is a ground pair and $v'$ is near $v$, then $\xs(v',\rho)$ is small.
Lemmas \ref{lem:excess-is-Lipschitz} and \ref{lem:all-excess-Lipschitz} give the
best forms of this claim.
We proceed to examine the situation in detail. 

In crudest terms, the next few lemmas are concerned with comparing functionals on
$\SV\times\SD$ and showing when one of them is large somewhere, then another one is also.
There are a lot of undetermined constants ($a$, $b$, $c$, etc.) in the statements,
and they cannot be assumed to have the same value from one occurence to the next,
except within a proof, as indicated by context. 
In the course of the demonstration that \ref{B:dom-E} is satisfied by the standard
interpretation (Section \ref{sec:RCI-demos}), 
Lemma \ref{lem:F-is-coercive} and part of Lemma \ref{lem:pair-is-o(F)} were effectively
already shown to hold in that interpretation. Now we will see that, conversely, they are
implied by the very simple \ref{B:dom-E}, with help from the other axioms.

\begin{lem}\label{lem:F-is-coercive}
  For some $a$ and $b$,
\begin{equation}
\|\rho\| \le a + bF(\rho).
\end{equation}
 \end{lem}
\begin{proof}
This is a consequence of local L-continuity of $E$ (Thm. \ref{prop:RCI}.2) as follows.
For some $r > 0$ and $c$, $E \ge c$ on the closed ball $\overline{B}(r)$
of radius $r$ about $v \equiv 0$.
Also, for any $\rho$, there is a $v \in \overline{B}(r)$ such that
\hbox{$\pair{v}{\rho} = -{r}\|\rho\|$} by the Hahn-Banach theorem.
Therefore,
\hbox{$F(\rho) \ge E(v) - \pair{v}{\rho} \ge c + {r} \|\rho\|$}.
\end{proof}
\begin{lem}\label{lem:pair-is-o(F)}
Given $\epsilon > 0$, each potential has a neighborhood $U$ such that
on $U\times\SD$
\begin{equation}
|\langle{v,\rho}\rangle| \le c(U) + \epsilon F(\rho).
\end{equation}
 \end{lem}
\begin{proof}
  We first prove that the bound holds for $v$ individually, and extend to a neighborhood
  afterward.
For any $M > 0$, and either choice in $\pm$, we have the inequality
\hbox{$F(\rho) \pm M \pair{v}{\rho} \ge E(\pm Mv)$}.
Together, they imply
\begin{equation}
  \nonumber
  \tfrac{E(Mv)}{M} - \tfrac{1}{M} F(\rho)
  \le \pair{v}{\rho} \le 
  -\tfrac{E(-Mv)}{M} + \tfrac{1}{M} F(\rho).
\end{equation}
Appealing to finiteness of $E(\pm Mv)$ (axiom \ref{B:dom-E}) and lower-boundedness of
$F$ (\ref{A:cvx}), this gives
\hbox{$|\langle{v,\rho}\rangle| \le c + \tfrac{1}{M} F(\rho)$} for some $c$.
Since $M$ may be taken as large as desired, this suffices.

Now, improve this to uniformity over $U$.
Let $\epsilon > 0$ be given.
By Lemma \ref{lem:F-is-coercive} the preceding paragraph,
\begin{alignat}{3}
  |\langle{v',\rho}\rangle| &
\le |\langle{v,\rho}\rangle| & + & |\langle{v'-v,\rho}\rangle|
\nonumber \\                              
& \le  c + \epsilon' F (\rho) & + & \|v'-v\|' ( a + b F(\rho) ).                        
  \nonumber
\end{alignat}
Here, $\epsilon'$ can be chosen as small as desired at the potential cost of large $c$.
Choose $\epsilon'$ and $r$ so that  $\epsilon' + b r < \epsilon$.
This ensures that whenever $\|v'-v\|' \le r$,
\hbox{$|\langle{v',\rho}\rangle| \le  (c + r a) + \epsilon F (\rho)$}.
\end{proof}
\begin{lem}\label{lem:F-and-excess-comparable}
Given $\epsilon > 0$, each potential has a neighborhood $U$, 
such that over \hbox{$U\times\SD$},
\begin{equation}
c + (1-\epsilon) F < \xs < c' + (1+\epsilon) F  
\end{equation}
\end{lem}
\begin{proof}
For any $v'$ and $\rho$,
\begin{equation}
  \nonumber
  |\xs(v',\rho) - F(\rho)| = |\pair{v'}{\rho} - E(v')|.  
\end{equation}
Choose $U$ such that both $E$ is bounded on $U$ (by Prop.~\ref{prop:RCI}.2)
and $|\langle{v',\rho}\rangle| < a + \epsilon F(\rho)$ for $v'\in U$
(by Lemma \ref{lem:pair-is-o(F)}).
\end{proof}

With the aid of the preceding lemmas, we turn to examining Lipschitz constants
for $v \mapsto \xs(v,\rho)$.

\begin{lem}\label{lem:excess-is-Lipschitz}
Every potential has a neighborhood $U$ on which the maps
\hbox{$v \mapsto \xs(v,\rho)$} are all locally \hbox{L-continuous} with local
Lipschitz constants $c + b F(\rho)$. Here, $c$ may depend on the neighborhood,
but $b$ does not.
\end{lem}
\begin{proof}
Using local L-continuity of $E$, choose a neighborhood $U$ on which
\hbox{$|E(v') - E({v})| \le L \|v' - {v}\|'$}. We confine our attention to
$U$ henceforth.
By definition of excess energy,
\hbox{$\xs({v},\rho)  - \xs({v}',\rho)
= \pair{{v}-v'}{\rho} + E(v') - E({v})$},
so \hbox{$|\xs({v},\rho) - \xs(v',\rho)| \le \Big(L+\|\rho\|\Big) \|v-v'\|'$}.
Now, apply Lemma \ref{lem:F-is-coercive} to bound $\|\rho\|$ here by
$c + b F(\rho)$.
\end{proof}
\begin{lem}\label{lem:all-excess-Lipschitz}
Any given potential has a neighborhood $U$ on which the maps
$v \mapsto \xs(v,\rho)$ are all L-continuous with
Lipschitz constants $c + b \inf_{v\in U}\xs(v,\rho)$.
\end{lem}
Here is a loose paraphrase. For $v$ varying over $U$,
$\xs(v,\rho)$ is either uniformly large, or does not vary much,
depending on $\rho$. In particular, the maps $v\mapsto \xs(v,\rho)$
for $\rho$'s which are ground densities of some potential in $U$ all
have a common Lipschitz constant over $U$.
\begin{proof}
Take $U$ to satisfy Lemma \ref{lem:excess-is-Lipschitz} and
Lemma \ref{lem:F-and-excess-comparable} with $\epsilon = 1$ (for instance).  
Lemma \ref{lem:excess-is-Lipschitz} implies that
\begin{equation}
  \nonumber
  |\xs({v}',\rho) - \xs(v'',\rho)| \le ( a + b\, F(\rho)) \|v'-v''\|'.
\end{equation}
Now apply Lemma \ref{lem:F-and-excess-comparable}.
\end{proof}

Finally, the preceding technical lemmas can be applied to obtain something
more digestible.
Recall that Prop. \ref{prop:RCI}.1 implies that if $(v_n,\rho_n)$ is a
sequence of ground pairs such that $v_n \to v$ and $\rho_n \to \rho$,
then $(v,\rho)$ is a ground pair. The next proposition shows that,
if the assumption that $(\rho_n)$ converges is dropped, we can still
assert that the $\rho_n$ are asymptotically nearly ground densities of $v$
in the sense of having small excess energy.
\begin{prop}\label{prop:asymptotically-small-excess}
  Given a sequence $(v_n,\rho_n)$ in \hbox{$\{\xs \le \epsilon\}$}
  such that $v_n \to v$, then $\limsup \xs(v,\rho_n) \le \epsilon$.
  More precisely,
\begin{equation}
\xs(v,\rho_n) \le \epsilon + c \|v-v_n\|'.
\end{equation}
\end{prop}
\begin{proof}
  If we restrict our attention to some tail of the sequence ($n \ge N$),
  all the $v_n$'s are in a neighborhood $U$ of $v$ satisfying
Lemma \ref{lem:all-excess-Lipschitz}.
Then, \hbox{$v'\mapsto \xs(v',\rho_n)$} is L-continuous over $U$
with Lipschitz constant $a+b\epsilon$, so that
\hbox{$\xs(v,\rho_n) \le (a +b\epsilon) \|v-v_n\|'$}.
The values $\xs(v,\rho_n)$ for $n < N$ can be accomodated in the same
kind of bound at the possible expense of increasing $a+b\epsilon$ to
some $c$.
\end{proof}

\section{Interlude: in pursuit of compactness}\label{interlude-compactness}

Prop.~\ref{prop:asymptotically-small-excess} demonstrates that,
when $(v_n,\rho_n)$ is a sequence of ground pairs with
$v\to \tgt{v}$, the situation is good with respect to excess energy.
The conclusion $\xs(\tgt{v},\rho_n)\to 0$ is similar to energetic progress
from section \ref{sec:progress}.
If we have a weaker metric than $d$, it will be easier for the sequence of
densities to converge, but the limit might not be a ground density of $\tgt{v}$
in that case. 
We turn our attention to finding a metric which is usefully weaker, but which
is strong enough that $\lim\rho_n$ will be a ground density of $\tgt{v}$.
We would be assured that the sequence at least had cluster points,
if we could guarantee that it was confined to a \emph{compact}, or
\emph{totally bounded} set.
That remark calls for a review of the important topological notion of
\textit{compactness}, in a form suitable for our purposes. 
Although no overt appeal to this concept is made until section \ref{sec:RCIII},
it already begins to exert an influence on the direction of the development.

\subsection{Compactness and total boundedness}

A helpful slogan is,
``a compact set is almost finite, in a topological sense''.
A metric space $X$ is said to be \textit{totally bounded} exactly if,
for any specified $\epsilon > 0$, there is
a finite set of points $x_1,\ldots,x_N\in X$ such that $X$ is covered by
the balls of radius $\epsilon$ centered at those points.
A complete, totally bounded metric space is \textit{compact}.
Although not the usual definition, this is equivalent to the latter,
and immediately captures the significance for our purposes.
If $(y_i : i\in \Nat)$ is \textit{any} (not necessarily Cauchy!)
sequence in a compact metric space $X$, then some subsequence converges to a
point in $X$. 

For example, any closed bounded interval $[a,b]\subset \Real$
($-\infty < a \le b < \infty$) is compact.
The entire real line is not, since the sequence $y_i = i$ has
no convergent subsequence. So, unboundedness is a way to avoid being
compact. Another is having infinitely many dimensions.
For instance, the closed unit ball of an infinite-dimensional Hilbert space is
not compact. If $\setof{\psi_i}{i\in\Nat}$ is an orthonormal basis, then,
the sequence $i \mapsto \psi_i$ does not converge in norm.
In an infinite dimensional Banach space,
a compact set is both bounded and ``almost finite-dimensional'' in being
within any prescribed distance of some finite-dimensional affine subspace.

\subsection{Total variation metric is a physically grounded candidate}

The weaker a metric on $\SD$, the more compact sets it will have.
We are thus motivated to consider metrics weaker than $d$, induced by the
norm $\|\cdot\|$.
Focusing on the standard interpretation, there is a particularly attractive
possibility, namely the metric $d_1$ induced by $L^1$ norm.

Earlier, we argued that, topologically, one should start from $\sigma(\SD,\SV)$.
If $\|\cdot\|_{1\cap 3}$ is physically motivated, then any metric strictly
between these two is also. This is not quite true of $d_1$.
As we shall see, $d_1$ is stronger than $\sigma(\SD,\SV)$ on $F$-bounded
sets, but not globally. However, $d_1$ has strong independent physical
credentials. 

First, the $L^1$ norm hews tightly to the very concept of density, as an
instrument for telling us how much ``stuff'' is in any specified region,
whereas the $L^3$ norm is, as observed, really a proxy for something else.
Indeed, if $\NN(A)$, respectively $\NN'(A)$, is the number of particles in region
$A$ according to density $\rho$, respectively $\rho'$, then
\begin{equation}
\|\rho - \rho'\|_1 = 2\cdot \sup \{ \NN(A)-\NN'(A)\,:\, A\subset \mathcal{M}\}.
\end{equation}
This metric has a privileged place in probability theory (a probability measure
taking the place of $\rho$), where it is known as \textit{total variation} metric.

Secondly, the map $\dens$ from quantum mechanical states (density matrices)
with the natural trace norm to $\vecsp\SD$ is continuous with respect to
total variation metric, but not $L^3$ norm. The former therefore has a direct
link to the underlying quantum mechanics as well.

The next section examines what it takes to replace $d$ by a weaker metric.
The motivation for this lies in the possibility of convenient compact sets,
but that theme will be put aside for now.

\section{Structure and regularity II}\label{sec:RCII}

This section is concerned with conditions (\ref{C:d1<d} -- \ref{C:clsc})
under which we can replace $d$ by a weaker metric, $d_1$ so that the product
space $\SV\times\SD$ continues to enjoy (nearly all of) the favorable properties
listed in Proposition~\ref{prop:RCI}, now with respect to the metric $d'+d_1$.
In the standard interpretation, $d_1$ is the $L^1$ metric.

\subsection{Complete lower semicontinuity}\label{sec:cmp-lsc}

In referring to a completion of $\SD$, axiom \ref{B:lsc} makes reference to
points outside $\SD$. With a weaker metric, we would have even more of these.
We would like to avoid that, due to the physically dubious status of those points,
and phrase everything in terms of $\SD$.
Here we identify the concept to do this, which turns out to be the same as appears
in item 1 of Prop.~\ref{prop:RCI}. Thus, we achieve some unification at the same time.

\begin{defn}[completely lower semicontinuous]
  A function $\Arr{(\SX,d)}{f}{\Real}$ on a metric space is
  \textit{completely lower semicontinuous} precisely if,
  for each $M < \infty$, the metric subspace $\{f\le M\} \subseteq \SX$ is complete
  (Def. \ref{defn:complete}).

  Normally, we are interested in a fixed function on the set $\SX$ and want
  to know whether $f$ is completely lower semicontinuous with respect to $d$.
  If so, we say that $d$ \textit{makes $f$ completely lsc}.
\end{defn}
Here is the fundamental fact about this concept.
\begin{lem}\label{lem:completely-lsc}
For $\Arr{(\SX,d)}{f}{\Real}$, these are equivalent:
  \begin{enumerate}[label=\alph*.]
  \item $\bar{f}$ defined on the completion $\overline{(\SX,d)}$ by
    \begin{equation}
      \bar{f}(x) \defeq
      \begin{cases}
        f(x) & x\in \SX \\
        +\infty & \text{otherwise}
      \end{cases}
    \end{equation}
is lower semicontinuous.
\item $f$ is completely lsc.
\item If $f$ is bounded above on the Cauchy sequence $(x_n)\subset (\SX,d)$,
  then it has a limit $x$ in $\SX$ and $f(x) \le \liminf_{n\to\infty} f(x_n)$.
\end{enumerate}
In particular, a completely lsc function is lsc.
\end{lem}
\begin{proof}
\textit{b} $\Leftrightarrow$ \textit{c} is elementary.
\medskip

\noindent\textit{a} $\Rightarrow$ \textit{c}:
  Assume \textit{a}, and let $(x_n)$ be a Cauchy sequence in $(\SX,d)$ on which
  $f$ is bounded by, say, $M < \infty$. It has a limit $x \in \overline{(\SX,d)}$,
  and by \textit{a}, $f(x) \le M$, and therefore $x\in\SX$.

\medskip
\noindent\textit{c} $\Rightarrow$ \textit{a}:
Let $(x_n)$ be a Cauchy sequence in $(\SX,d)$ with limit
\hbox{$x\in \overline{(\SX,d)}$}, such that
\hbox{$\liminf f(x_n) < \infty$}. (If this condition fails, there is nothing to show).
Take \hbox{$c > \liminf f(x_n)$}.
Then, the subsequence consisting of $x_m$ for which
{$f(x_m) < c$} is bounded above and also converges to $x$.
Apply (c) to this subsequence to conclude $f(x) \le \liminf f(x_n)$.
\end{proof}

\subsection{New postulates}

$d_1$ is a metric on $\SD$ such that
\begin{enumerate}[label=C{\arabic*}.,series=caxs,ref=C\arabic*]
\item\label{C:d1<d}
  $d_1 \precsim d$ 
\item \label{C:F-good}
$\sigma(\SD,\SV) \precsim d_1$ on $F$-bounded sets
\item \label{C:clsc}
  $d_1$ makes $F$ completely lsc
\end{enumerate}

By \ref{B:lsc} and Lemma \ref{lem:completely-lsc}, the metric $d$ itself
satisfies these axioms.
Of course, we have in mind a different, strictly weaker candidate
for $d_1$, namely the $L^1$ metric, motivated by compactness properties
to be discussed in later sections.
These new axioms actually render \ref{B:lsc} redundant, because the
properties in \ref{C:F-good} and \ref{C:clsc} are stable under
strengthening $d_1$.
\begin{prop}
Given \ref{A:D} -- \ref{A:dual-pair},
\ref{C:d1<d} -- \ref{C:clsc} imply \ref{B:lsc}.  
\end{prop}

\subsection{Standard interpretation}

The new ingredient here is $d_1$.
The standard interpretation is that $d_1$ is the metric induced by
the $L^1$ norm $\|\cdot\|_1$, as discussed in the Interlude.
This is the motivation for the subscript $1$ on `$d_1$'.

\subsection{Improved structure theorem}

From now on, we assume
\ref{A:D} -- \ref{A:dual-pair},\ref{B:dom-E}, \ref{B:norm}, \ref{C:d1<d} -- \ref{C:clsc}.

The metric on $\SD$ is $d_1$, and this will continue to be the metric of interest
in following sections. On the other hand, we continue to use the metric $d'$ on $\SV$.
Note that even if $d_1$ comes from a norm (which we do not require),
$\SV$ is generally smaller than the dual of $(\vecsp\SD,d_1)$.
The statement of the main proposition in this section is similar to
that of Prop.~\ref{prop:RCI}, but for the use of the new terminology.
The use of $d_1$ instead of $d$ indicates that
the proposition is stronger than Prop.~\ref{prop:RCI}, except for the minor point
that we no longer obtain Lipschitz continuity of $F$ on $\zero$.
The proof is given in Section \ref{sec:RCII-proof}.
\begin{prop}\label{prop:RCII}
On $(\SV\times\SD,d'+d_1)$, 
\begin{enumerate}
\item $\xs$ is completely lsc
\item $E$ is locally L-continuous
\item $F$ is $\xs$-almost continuous
\item $F$ is continuous on $\zero$
\end{enumerate}
\end{prop}
\subsection{\ref{C:d1<d} -- \ref{C:clsc} hold in standard interpretation}
\label{sec:RCII-demo}

  
\begin{proof}[Proof of \ref{C:d1<d}]
Immediate from definition.  
\end{proof}

\begin{proof}[Proof of \ref{C:F-good}]
  Assume $(\rho_n) \subseteq \{F\le M\}$ converges to \hbox{$\rho \in \{F \le M\}$}
  with respect to $\|\cdot\|_1$.
We need to show the convergence holds also with respect to $\sigma(\SD,\SV)$.
Lemma \ref{lem:F-is-coercive} gives a bound $\|\rho_n-\rho\|_3 < c$,
while Lemma \ref{lem:v-is-nearly-L-infty} allows to write $v = v'+v''$,
with $v'$ bounded and $\|v''\|_{3/2} < \epsilon/c$.
So, $\pair{v_1}{\rho_n-\rho} \to 0$ by assumption, while
$|\langle{v_2,\rho_n - \rho}\rangle| < \epsilon$ for all $n$.
That is, \hbox{$\liminf |\langle{v,\rho_n - \rho}\rangle| < \epsilon$}.
Since $\epsilon > 0$ is arbitrary, that is what is needed.
\end{proof}

\begin{proof}[Proof of \ref{C:clsc}]
This is a consequence of lower semicontinuity of $F$ as a
function on $L^1(\Real^3)$ when extended as $+\infty$ off $\SD$.
See Thm. 4.4 of Ref. \onlinecite{Lieb83} for details of the latter.
\end{proof}

\subsection{Proof of Proposition \ref{prop:RCII}}\label{sec:RCII-proof}

\noindent A. {$E$ is locally L-continuous}.
\begin{proof}
$E$ depends only on $v$, and the norm on $\SV$ has not changed,
so this is the same as in section~\ref{sec:RCI-proof}.
\end{proof}

\medskip\noindent B.
$d'+d_1$ makes $\xs$ completely lsc on $\{F\le M\}$.
\begin{itemize}
\item 
  If $(\rho_n) \subset \{F\le M\}$ is
  $d_1$-Cauchy, then it is \hbox{$\|\cdot\|$-bounded}. (old norm here!) \newline
  \textit{Proof}:
$\{\rho_n\}$ is $d_1$ bounded because $d_1$-Cauchy,
hence also $\sigma(\SD,\SV)$-bounded by \ref{C:F-good}.
Therefore, by the Uniform Boundedness Principle, $\{ \|\rho_n\| \}$ is bounded.

\item $(v,\rho) \mapsto \pair{v}{\rho}$ is continuous on $\{F\le M\}$. \newline
\textit{Proof}:
Suppose $(v_n,\rho_n) \xrightarrow{d' + d_1} (v,\rho)$.
We need to show that \hbox{$\pair{v_n}{\rho_n}\to \pair{v}{\rho}$}.
Now,  
\begin{equation}
  \nonumber
\pair{v}{\rho} - \pair{v_n}{\rho_n} = \pair{v}{\rho-\rho_n} + \pair{v-v_n}{\rho_n}.
\end{equation}
Check that each term on RHS tends to zero:
\newline
1st term: By hypothesis on $d_1$,
$\rho_n \xrightarrow{\sigma(\SD,\SV)} \rho$.
That does it, since the pairing is with fixed $v$.
\newline
2nd term:
$|\pair{v-v_n}{\rho_n}| \le \|{v-v_n}\|' \|{\rho_n}\|$,
and $\|{v-v_n}\|' \to 0$, while $\|\rho_n\|$ is bounded by the preceding bullet point.
\item
$\xs$ is therefore completely lsc on $\{F \le M\}$ because it is the sum
of a completely lsc function ($F$, by axiom \ref{C:clsc}) and two continuous functions,
namely $E$ (by A) and \hbox{$(v,\rho)\mapsto \pair{v}{\rho}$} (by preceding bullet).
\end{itemize}

\medskip\noindent C. $\xs$ is completely lsc.
\newline
Now we lift the restriction to $\{F\le M\}$.
Suppose $((v_n,\rho_n)) \subset \{\xs\le \epsilon\}$ is a Cauchy sequence
with $v_n \to v$. Some tail of the sequence is in the neighborhood $U$ of
Prop.~\ref{lem:F-and-excess-comparable}, hence is $F$-bounded.
So, without assuming $F$-boundedness, we get it anyway, and recover the
situation of item B.

\medskip\noindent D. On $\{\xs\le\epsilon\}$, $F$ is $\epsilon$-almost continous.

\begin{proof}
The proof is formally just like that for item E in section~\ref{sec:RCI-proof}.
\end{proof}

\medskip\noindent E. $F$ is continuous on  $\zero$.

\begin{proof}
Special case of item D.
\end{proof}

\section{Density clustering and tightness}\label{sec:RCIII}

This Section assumes
\ref{A:D}--\ref{A:dual-pair}, \ref{B:dom-E},\ref{B:norm},\ref{C:d1<d} -- \ref{C:clsc}.

We aim for a simple criterion to guarantee that whenever $(v_n,\rho_n)$
is a sequence of ground pairs and $v_n \to v$, then $\rho_n\to \rho$ with
respect to $d_1$. That would
ensure that $\rho$ is a ground density of $v$, by Prop.~\ref{prop:RCII}.1.
Actually, this is asking too much. Instead of asking for convergence of
the sequence $(\rho_n)$, we ask only that it \textit{cluster} on a nonempty set $D$.
This means that every density in $D$ is the limit of a subsequence of $(\rho_n)$,
and every subsequence of $(\rho_n)$ has a further subsequence converging to something
in $D$. An alternative way to say the same thing is the following.
Denote the $d_1$ closure of $\{\rho_m\}_{m\ge n}$ by $T_n$.
This is a decreasing (with $n$) sequence of closed sets, and the equivalent
statement is that the limit (i.e., intersection $\cap_n T_n$) is precisely $D$.
The important thing is that, every density in $D$ is a ground density for $v$,
in this case.

\subsection{A first attempt}

We begin our search for a criterion with
\begin{lem}\label{lem:silly}
Let $(v_n,\rho_n) \subseteq \{\xs \le \epsilon \}$ be a sequence such that $v_n \to v$.
Then, $\{\rho_n\}$ is $F$-bounded.
If, in addition, $\{\rho_n\}$ is $d_1$-totally bounded,
then, relative to $d_1$, the sequence $(\rho_n)$ clusters on a set $D$,
 such that \hbox{$(v,\rho) \in \{\xs \le \epsilon\}$} for every $\rho\in D$.
\end{lem}
\begin{proof}
  That $\{\rho_n\}$ is $F$-bounded is a simple consequence of
Lemma \ref{lem:F-and-excess-comparable}: For some $N$,
$\setof{v_n}{n\ge N}$ is in the neighborhood $U$ of that Lemma.
Since $\setof{(v_n,\rho_n)}{n \ge N}$ is $\xs$-bounded,
$\setof{\rho_n}{n \ge N}$ is $F$-bounded. 

Thus, for some $M < \infty$, $\{\rho_n\} \subseteq \{F \le M\}$,
which is a complete metric space under $d_1$ by \ref{C:F-good}.3.
By hypothesis, $\{\rho_n\}$ is $d_1$-totally bounded, so its closure in
$\{F \le M\}$ is compact.
Therefore, there is a set $D$ of densities such that $(\rho_n)$ clusters on $D$.
But, now we are dealing with sequences such that both components, $v_n$ and
$\rho_n$, converge. Prop. \ref{prop:RCII}.1 completes the proof.
\end{proof}
So, the density components cluster on ground densities of $v$ if the sequence
of densities is \hbox{$d_1$-totally} bounded, and one is
tempted to consider this condition to be the answer to our problem.
Physically, though, it is not very simple or transparent. We will keep looking. 

The first part of the Lemma says that the set $\{\rho_n\}$ is necessarily
$F$-bounded. Therefore, what we should look for is a property of
sets of densities which guarantees that it is $d_1$-totally bounded
as soon as it is $F$-bounded. For brevity, we will call such a property a
\textit{compactness test}. This is too special to be enshrined as an official
definition. So, to repeat: $\mathscr{P}$ is a compactness test if every $F$-bounded
set which is $\mathscr{P}$ (we use it as an adjective) is $d_1$-totally bounded.
\begin{prop}\label{prop:compactness-test}
Let $\mathscr{P}$ be a compactness test, 
and $(v_n,\rho_n) \subseteq \{\xs \le \epsilon \}$ a sequence such that $v_n \to v$.
Then, if $\{\rho_n\}$ is $\mathscr{P}$, the sequence $(\rho_n)$ clusters on a
set of ground densities for $v$.
\end{prop}
\begin{proof}
Follows immediately from the definition of compactness test and Lemma \ref{lem:silly}.  
\end{proof}

The axioms do not seem very helpful in finding a compactness test, so we
will look more closely at the special features of the standard
interpretation.

\subsection{Tightness}\label{sec:tightness}

Let us approach the problem from a different angle.
In standard interpretation, if the sequence of densities $(\rho_n)$ is to converge,
it is certainly necessary that the following hold:
given arbitrary $\epsilon > 0$, there is some sphere such that,
from some point in the sequence on, $\rho_n$ puts less than
particle number $\epsilon$ outside the sphere.
Otherwise the sequence is ``leaky'' or ``lossy'' in the sense that
some nonzero particle number is inexorably moving off to infinity.
This \textit{necessary} condition is called \textit{tightness}.
It is also sufficient. In conjunction with $F$-boundedness, guaranteed
by Lemma \ref{lem:silly}, tightness implies $d_1$-total boundedness.
We pass to details.

\begin{defn}[tight]
A set $\mathcal{F}$ of integrable functions on $\Real^d$ is \textit{tight} if,
for every $\epsilon > 0$, there exists $R$ such that
for every $f \in \mathcal{F}$,
\begin{equation}
  \label{eq:tight}
\int_{|x| > R} |f(x)|\, dx < \epsilon.
\end{equation}
\end{defn}
In using this notion, we are implicitly working in an interpretation, in particular
of $\SD$, in which it makes sense.
Tightness seems to be a property more easily reasoned about than $d_1$-total
boundedness. It is especially so if we are content with just a fair
level of confidence, since a lot of quantum mechanical intuition can be brought
to bear on it.

\begin{lem}\label{lem:tight-F-bdd-tot-bdd}
  $F$-bounded tight subsets of $\SD$ are \hbox{$d_1$-totally} bounded.
  In other words, tightness is a compactness test in the standard interpretation.
\end{lem}
\begin{proof}
There are four ingredients.
  
\noindent 1. The Rellich-Kondrachov theorem, a standard tool in theory of Sobolev spaces.
  (See, for example, Thm. 9.16 of Ref. \cite{Brezis}.)
  For our purposes, it says: If $\Omega$ be a bounded subset of $\Real^n$,
  and $K \subset L^1(\Omega)$ such that both $\|f\|_1$ and $\|\nabla f\|_1$
  are bounded over $K$, then $K$ is totally bounded in $L^1(\Omega)$.

\noindent 2. If $A$ is a set of densities in $\{F \le M\}$, then
$\|\nabla \rho\|_1$ is bounded over $A$. See (4.7) of Ref. \cite{Lammert14}.

\noindent 3. To apply the Rellich-Kondrachov theorem, we need to be able to
  ignore the tails of the densities. This is the role of tightness.
  The general principle is this. A set $A$ in a metric space is totally
  bounded if, for any $\epsilon > 0$, there is a totally bounded set $K$
  such that $A$ is in the $\epsilon$-dilation of $K$ (every point of $A$ within
  $\epsilon$ of $K$). Using this, given $\epsilon$, take $\Omega$ to be
  the ball $B(R)$ with $R$ as in (\ref{eq:tight}).

  \noindent  4. $L^1(\Omega)$ is isometrically embedded in $L^1(\Real^n)$, so that
  a totally bounded subset of the former can be construed as a totally bounded
  subset of the latter.
\end{proof}
This result has a nice semiclassical interpretation.
The idea is that a volume $h^{\NN}$ in phase space corresponds
to one dimension in Hilbert space.
Now, if $\mathcal{F}$ is tight, then densities in $\mathcal{F}$ come from states almost
bounded in position, and the bound on $F$ implies a bound on momentum.
This gives us that $\dens^{-1}(A\cap\{F\le M\})$ is an ``almost finite dimensional''
set of density matrices, i.e., it is compact.
Since the map $\Arr{\mathcal{L}_1(\HH)}{\dens}{L^1(\Real^3)}$ is
continuous, the image of that compact set is compact.

Finally, combining Lemma \ref{lem:tight-F-bdd-tot-bdd} and Lemma~\ref{lem:silly},
we reach the objective of this section, and a major objective of the paper.
\begin{prop}\label{prop:auto-density}
 (standard interpretion)
If \hbox{$(v_n,\rho_n) \subset \{\xs \le \epsilon\}$}, $v_n\to v$,
and $(\rho_n)$ is tight, then it clusters on a set of densities $D$
such that $\xs(v,\rho) \le \epsilon$ for every $\rho\in D$.
\end{prop}

\subsection{Interpretations with automatic tightness}

There are at least a couple of interesting variations on the standard
interpretation in which tightness is automatic, and therefore we require
\textit{no} condition on the sequence $(\rho_n)$ in the above setting.
One such is the case where
$\mathcal{M}$ is not $\Real^3$, but a three-torus, or more generally
a closed manifold. In that case,
$L^1(\mathcal{M})$ is isomorphic to $L^1([0,1]^3)$.
If it is thought of that way, all sequences in $\SD$ are tight.

Another case leaves everything as in the standard interpretation,
except $F$, which is further specialized (beyond what the axioms say)
to an exact functional with a repulsive interaction, and a background trap
potential tending to $+\infty$ as $|x| \to \infty$.
For example, a harmonic potential.
In this case, the condition (\ref{eq:tight}) is implied by $F$-boundedness.

%
%
%
%
%
%
\section{Recapitulation}

Here is a very brief, and necessarily imprecise, recapitulation
of the findings, with emphasis on the standard interpretation and exact
functionals, hence cutting out the axiomatic middlemen.

Kohn-Sham computation can be viewed as a walk on ground pairs in $\SV\times\SD$.
Indeed, the entire development is based on a commitment to think, explicitly,
in this bivariate way.
A simple iterative scheme, focusing on potential, is shown to make progress
(with caveats) in the sense of being able to move to a density with lower
excess energy $\xs(\tgt{v},\rho)$ in the presence of the target potential.
Somewhat surprisingly, no metrics on potential or density space is required
to carry out that analysis.
For a deeper treatment, in particular to discuss convergence questions,
however, some metric or topological structure is necessary.
With respect to the metric $d'+d_1$ on $\SV\times\SD$, the
following hold: Ground energy $E$ is continuous, while intrinsic energy
$F$ and excess energy $\xs$ are completely lower semicontinuous.
$F$ is also $\xs$-almost continuous. Thus, although $F$ is unbounded
above on every neighborhood, this phenomenon and possible unpleasant
consequences are strongly mitigated as long as we restrict attention to the
low intrinsic energy subspace, and $F$ is even continuous on $\zero$.
Low excess energy pairs are close
to the set $\zero$ of ground pairs, metrically.
Conversely, $\xs$ increases only slightly when shifting the potential
of a point in $\zero$. (The corresponding statement with respect to density
is absolutely \emph{not} true, not even for the $\|\cdot\|$ metric.)
If $(v_n,\rho_n)$ of ground pairs is such that $v_n \to \tgt{v}$, then
the densities automatically accumulate on ground densities of $\tgt{v}$,
as long as the density sequence does not have particle number drifting to
infinity.

\section{Some conclusions}

This work aimed to bringing rigorous mathematical analysis of DFT a little
closer to the computational practice of DFT, and in the process to
get a more physical picture of both.
It is based on a few simple ideas.
First, the procedures and operations of KS computation should be physically
interpreted.
Second, the topologies (norms) on potential and density spaces entering a
functional analytic theory also require physical grounding.
Third, one should work explicitly in the product of potential and density
space as much as possible.
These are also, especially the last, conclusions as starting points.
They are vindicated by the results achieved in taking them seriously.

A number of the results in this paper point to the somewhat ironic
conclusion that more attention should be payed to potential in density functional theory.
These are, primarily, the demonstration in section \ref{sec:feasible-strategy}
that an iterative scheme focusing on potential can make progress, with provisos,
and the result, Prop.~\ref{prop:auto-density}, on automatic convergence of density. 


\begin{acknowledgments}
This project was funded by the National Science Foundation under award DMR-2011839.
The author is grateful to a Referee who pointed out a circular argument in a
previous version.
\end{acknowledgments}
%


\end{document}